\documentstyle{l-aa}              
\input epsfig.sty

\def\simless{\mathbin{\lower 1pt\hbox
   {$\spose{\raise 5pt\hbox{$\char'074$}}\char'430$}}}
\def\simgreat{\mathbin{\lower 1pt\hbox
   {$\spose{\raise 5pt\hbox{$\char'076$}}\char'430$}}}
\def\lapp{\mathbin{\raise2pt \hbox{$<$} \hskip-9pt \lower4pt \hbox{$\sim$}}}
\def\gapp{\mathbin{\raise2pt \hbox{$>$} \hskip-9pt \lower4pt \hbox{$\sim$}}}
\begin{document}

   \thesaurus{07         
              (02.13.2;  
               06.19.2;  
               08.01.3;  
               08.13.2;  
               08.16.5;  
               11.10.1;  
               09.10.1)} 
   \title{Nonradial and nonpolytropic astrophysical outflows\protect\\
         IV. Magnetic or thermal collimation of winds into jets?}

  \author{C. Sauty
           \inst{1}
   \and    K. Tsinganos
           \inst{2}
   \and    E. Trussoni
           \inst{3}
          }
 
   \offprints{C. Sauty, sauty@obspm.fr}
 
   \institute
         {Observatoire de Paris -- Universit\'e Paris 7, 
          DAEC, F-92190 Meudon, France
    \and  Department of Physics, University of Crete and FORTH,   
         P.O. Box 2208, GR-710 03 Heraklion, Crete, Greece
    \and Osservatorio Astronomico di Torino, Strada
         dell'Osservatorio 20, I-10025 Pino Torinese (TO), Italy
         }
 
   \date{Received 9 November 1998 / Accepted 31 May 1999}
 
   \maketitle
   \markboth{Sauty et al. 
             Nonradial and nonpolytropic astrophysical outflows. IV.}{}
   \begin{abstract}

An axisymmetric MHD model is examined analytically to illustrate some key
aspects of the physics of hot and magnetized outflows which originate in the
near environment of a central gravitating body. By analyzing the
asymptotical behaviour of the outflows it is found that they attain  a
variety of shapes such as conical, paraboloidal or cylindrical. However, non
cylindrical asymptotics can be achieved only when the magnetic pinching is 
negligible and the outflow is overpressured on its symmetry axis. 
In cylindrical jet-type asymptotics, the outflowing plasma
reaches an equilibrium wherein it is confined by magnetic forces or 
gas pressure gradients, while it is supported by centrifugal forces or 
 gas pressure gradients. In which of the two regimes (with thermal or magnetic 
confinement) 
a jet can be found depends on the efficiency of the central magnetic rotator.
The radius and terminal speed of the jet are analytically given in terms of the
variation across the poloidal streamlines of the total energy. Large radius of
the jet and efficient acceleration are best obtained when the external 
confinement is provided with comparable contributions by magnetic pinching and
thermal pressure. In most cases,  collimated streamlines undergo oscillations 
with  various
wavelengths, as also found by other analytical models. Scenarios for the 
evolution of outflows into winds and jets in the different confinement regimes
are shortly outlined. 

\vskip 0.5 true cm

   \end{abstract}
   

\section{Introduction}

Nonuniform plasma outflows seem to be ubiquitous in astrophysics  on galactic
and extragalactic scales.  The closest example is the solar wind itself 
which shows strong heliolatitudinal velocity gradients as recently observed by 
{\it Ulysses} (Lima \& Tsinganos 1996, McComas et al. 1998). Further away 
collimated outflows are observed in association with several galactic objects, 
such 
as young and evolved stars, planetary nebulae, X-ray binaries and collapsed
objects (for reviews see Ray 1996, Kafatos 1996, Mirabel \& Rodriguez 1996, 
Brinkmann \& M\"uller 1998, Livio 1998). Finally, on extragalactic scales 
jets are observed to originate in many Active Galactic Nuclei and Quasars 
(Biretta 1996, Ferrari et al. 1996).

Yet, despite their abundance the basic questions on the formation, 
acceleration and propagation of nonuniform winds and jets have not 
been fully answered. Nevertheless, observations seem to indicate  
that the basic ingredients for producing astrophysical outflows are  
some sort of heating to launch thermally the wind at the axis plus a 
rotating central gravitating object and/or an accretion disk threaded 
with magnetic fields to 
accelerate magnetocentrifugally and collimate the outflow. 

\subsection{Drivers of the collimated plasma outflow}
Several mechanisms have been investigated for accelerating and collimating 
astrophysical outflows in galactic and extragalactic scales.  
Magnetic rotator forces seem to play
a rather dominant and crucial role (Lynden-Bell 1996) but they are probably
not the only relevant mechanism.

\noindent
{\it First}, thermally  driven models are based on the {\it de Laval
nozzle} analogy of the solar wind (Parker 1963, Liffman \& Siora 1997).  
This requires the presence of a hot corona around the central body of the 
YSO or the AGN. 
X-ray emission detected in several of these objects may imply that
thermal effects contribute to the general acceleration mechanism at the
base of the flow but they are probably not the only ingredient. Furthermore,
if the wind is associated to a very bright object, the flow can be effectively
accelerated by the photon flux (radiatively driven winds, Cassinelli 1979).
Parallely note also that collimation of bipolar outflows from YSOs and 
Planetary Nebulae by external thermal pressure gradients have been extensively
studied in the frame of the Generalized Wind Blown Bubble scenario 
(GWBB, Frank 1998). It has demonstrated successfully that magnetic processes 
may not be the only way to achieve collimated outflows. 
 
\noindent
{\it Second}, magnetic pressure driven models are based on the {\it uncoiling
spring} analogy and have been examined by Draine (1983), Uchida \& Shibata
(1985) and Contopoulos (1995). There, it is assumed that a toroidal magnetic
field $B_{\phi}$ is created and highly amplified by the winding-up of its field
lines by a radially collapsing and non-Keplerian rotating disk. Plasma is then
accelerated from the disk in the poloidal direction by the action of the
resulting torsional Alfv\'en waves. 

\noindent
{\it Third}, magnetocentrifugally driven outflow models are based on the
classical {\it bead on a rotating rigid wire} analogy. There, the magnetized
cold fluid is flung out (even to relativistic velocities) from the surface of
the Keplerian accretion disk, provided that the poloidal field lines are
inclined enough with respect to  the disk axis (Blandford \& Payne 1982,
Pelletier \& Pudritz 1992, Contopoulos \& Lovelace 1994, Cao  1997). This
approach is suitable to model winds from accretion disks, but is not valid
around the  symmetry axis. 
Moreover it has been pointed out recently (Ogilvie \& Livio 1998) that, even if
the lines are sufficiently inclined, a potential barrier still exists that can 
be overcome only by the presence of an extra source of energy (e.g.  a hot 
corona).

In all the above treatments the effects of the {\it combination} of gas 
pressure
\underbar{and} magnetic fields in accelerating, collimating and confining jets
have not been discussed adequately, despite the fact that the role of the gas
pressure has been recognized for a long time, i.e., that jets are not moving in
a vacuum (molecular clouds around YSO's, or host galaxies in AGN) and hence
they must have some interaction with the external medium (Ferrari et al. 1996;
Frank 1998). 
This approach may also highlight the transition from  fully
thermally driven to fully  magnetocentrifugally driven collimated winds. 

\subsection{Radially self-similar models}
As with any
fully MHD approach and despite of the simplifications of steadiness and
axisymmetric geometry, several approximations are still unavoidable in order to
obtain exact solutions useful for an understanding of the MHD mechanism for the
initial acceleration and final collimation. Thus, one simple analytical way out
is the use of self-similarity. This hypothesis allows an analysis in a 2-D
geometry of the MHD equations which reduce then to a system of ordinary
differential equations. The basis of the self-similarity treatment is the
assumption of a scaling law of one of the variables as function of one of the
coordinates. The choice of the scaling variable depends on the specific
astrophysical problem. 

Several models self-similar in the {\it radial} direction have been 
investigated to
analyze the structure of winds from accretion disks (Blandford \& Payne 1982,
Contopoulos \& Lovelace 1994, Li et al. 1992, Li 1995, 1996, Ferreira 1997,
Ostriker 1997). In these models 
the driving force and the collimation derive from a combination
of the magnetic and centrifugal forces. Moreover, as disc-winds are associated
with jets, these studies usually do not consider under which parametric
conditions full collimation is obtained. Exceptions are given in Pelletier \&
Pudritz (1992)  and Contopoulos \& Lovelace (1994) where the collimation
efficiency is linked to a current flowing in agreement with the Heyvaerts \&
Norman (1989) general analysis. However, the absence of an exact crossing of all
the existing critical points in the solutions presented in these papers
prevents from considering their conclusions as definitive. Nevertheless, it 
has been
shown that within the frame of self-similar disc-wind assumptions, it is
possible to cross all critical points thus getting meaningful solutions
(Tsinganos et al. 1996, Vlahakis 1998).  Moreover, the role of the
inhomogeneity in the pressure distribution has not been taken into account
until recently in these models (Ferreira 1997) and a full parametric study of
this extra variable is yet to be performed. 

\subsection{Meridionally self-similar models}

In a series of studies, solutions of the MHD equations that are self-similar 
in the {\it meridional} direction have been also analyzed (Tsinganos \& 
Trussoni 1990, 1991, 
Tsinganos \& Sauty 1992a,b, Papers I and II of this series,
Trussoni \& Tsinganos 1993,  Sauty \& Tsinganos 1994, Paper III of this series,
Trussoni et al. 1997, henceforth TTS97). Such a treatment 
allows to study the physical
properties of the outflow close to its rotational axis. As in this region the
contribution to acceleration of the magnetocentrifugal forces is small, the
effect of a thermal driving force is essential. This implies also that the 
structure of the gas pressure in the flow is essential. 

Two main classes of such self similar solutions have been investigated 
depending on whether the
components of the pressure gradient along the radial and meridional directions
are related or not. In the second case the shape of the streamlines and
fieldlines is prescribed `a priori', and the main features of the dynamical
variables are self-consistently deduced from the integration. 
In particular, it has been shown
that acceptable solutions for magnetized flows with asymptotic superAlfv\'enic
velocity exist only when rotation is included (Tsinganos \& Trussoni 1991,
Trussoni \& Tsinganos 1993, TTS97). As a consequence of this study it seems
that even pressure confined jets from slow magnetic rotators need magnetic
fields and rotation. 

In the other case, in which the two components of the gas pressure are related,
the structure of the streamlines is deduced as a self-consistent solution of
the MHD equations. 
It has been shown (Papers I and II) that
hydrodynamical and nonrotating magnetized winds are always radially expanding
from the source. On the other hand, rotating magnetized flows with a 
spherically symmetric structure for the pressure gradient can have final 
superAlfv\'enic velocities with either radial or collimated asymptotically 
streamlines, depending on the values of the parameters (Paper III). 
 This allows 
to deduce a criterium to select conically expanding winds from cylindrically
collimated jets (Sauty et al. 1996). 

\subsection{Plan of this paper}

We extend here the analysis of Paper III to the
more general case of solutions for rotating magnetized winds with a {\it
nonspherically} symmetric gas pressure. In the present paper we concentrate
on the asymptotic analysis and its link to the initial boundary conditions:
this allows us to derive a general criterion for the collimation of winds 
into jets. The analysis of the properties of the complete numerical solutions 
deserves a separate  study which is postponed to a following paper.  

In Sec. 2 we summarize the properties of the meridional self-similar MHD
equations  while in Sec. 3 we discuss the energetic structure of the outflow.
In particular we show that an energy integral exists that links the asymptotic
regime to the boundary conditions at the base, allowing to formulate a general
criterion for the collimation of the wind. The different physical conditions for
asymptotic confinement (magnetic or thermal) are discussed in detail in Sec. 4,
and in Sec. 5 we show that oscillating configurations can be present in
cylindrically collimated jets. In Sec. 6 the equilibrium asymptotic properties
of non collimated flows are outlined, while in Sec. 7 we summarize the
results and shortly discuss the astrophysical implications of our analysis.


\section{Meridionally self-similar MHD model}

\subsection{Steady axisymmetric ideal MHD outflows}

The global dynamical properties of cosmic winds and jets are usually analyzed
by assuming that they represent outflows of a fully ionized plasma with a
bulk speed $\vec{V}$ and carrying a magnetic field $\vec{B}$ in the 
gravitational field of a central body of mass $\cal M$. The familiar MHD
equations are employed for a physical description of these phenomena. In
particular, under steady and axisymmetric conditions ($\partial /\partial t =
\partial /\partial \varphi =0$), the MHD equations are known to admit certain
free integrals, i.e., functions which remain constant on the magnetic surfaces
generated by the revolution around the magnetic/flow symmetry axis of the
system of a poloidal magnetic line $A(r, \theta )$ = constant 
(Tsinganos 1982). Specifically, on the surface of such a flux tube $A$ = const.,
the following physical quantities remain invariant throughout the extent of
these surfaces from the base to infinity: 

\begin{itemize}
\item{$\Psi_A$(A)}, the ratio of the magnetic and mass fluxes, 
\item{$L(A)$}, the {\it total} specific angular momentum carried by the 
flow {\it and} the magnetic field,
\item{$\Omega(A)$}, the corotation frequency or angular velocity 
of each streamline at the base of the flow. 
\end{itemize}

Furthermore, it is well known that the poloidal ($p$) and azimuthal 
(toroidal, $\varphi$) components of the magnetic field and the 
velocity can be expressed in terms of these free integrals and the 
poloidal Alfv\'en Mach number, using spherical ($r, \theta, \varphi$) 
or cylindrical ($\varpi, \varphi, z$) coordinates (for details see 
Paper III). In particular, the poloidal Alfv\'en Mach 
number (or Alfv\'en number) $M$ is,
$$
M^2 = 4 \pi \rho {{V^2_p} \over {B^2_p}} = {{\Psi^2_A} \over {4 \pi 
\rho}}
\,.
\eqno(2.1)
$$
\noindent
On the other hand, the two integrals $L(A)$ and $\Omega(A)$ are not independent
{\it if} the flow is transalfv\'enic. In such a case, at the cylindrical distance
$\varpi_a$ of the Alfv\'en point ($M=1$) from the field/flow axis of a flux
tube labeled by $A$ they are related as $\varpi_a^2 (A) = ({r_* \, {\sin}
\theta_a})^2=L/\Omega$. 

\subsection{Generalized Bernoulli integral}

A fourth constant of the motion expresses the conservation of energy along 
streamlines. Thus, by projecting the momentum equation along a streamline,
taking into account the first law of thermodynamics for energy
conservation, we obtain the generalized classical Bernoulli integral
(Paper III),
$$
E(A)= {1\over2}V_{\rm p}^2 + {1\over2}V_{\varphi}^2 
- {{\cal G}{\cal M}  \over r} - {\Omega \, \over \Psi_A} \varpi B_\varphi
+ h - \Theta_{r_o}^r(A)  
\,,
\eqno (2.2\hbox{a})
$$
where
$$
\Theta_{r_o}^r(A) = \int_{r_o}^{r}{q(r',A)\over\rho(r',A)V_r (r',A)} {\rm d}r'
\,,
\eqno (2.2\hbox{b})
$$
 $h$ is the enthalpy of the perfect monoatomic gas ($\Gamma=5/3$), 
$q$ is the net local volumetric heating/cooling rate, and ${\cal G}$ 
the gravitational constant. Thus, at a given radial distance $r$ along 
the streamline labeled by $A$, the conserved energy $E(A)$ represents 
the sum of the kinetic, gravitational, Poynting and ideal thermal energy 
flux densities per unit of mass flux density, minus the extra heat 
received by the flow between the anchored footpoint at a basal radial 
distance $r_o$ and the point $r$ under consideration, $\Theta_{r_o}^r(A)$.

\subsection{Self-similarity: scaling laws for the variables}

The model analysed in this paper belongs to the wide class of  
meridionally self-similar MHD equilibria (see also Trussoni et al. 1996; 
Tsinganos et al. 1996; TTS97; Vlahakis \& Tsinganos 1998, henceforth VT98). 
In the following we briefly summarize 
the main steps for the construction of such a model (see Appendix A for 
more technical details).  

For convenience, first of all the variables are normalized to their 
respective values at the Alfv\'en surface along the axis of rotation, $r=r_*$. 
In particular, we define the dimensionless radial distance $R=r/r_*$ 
and the Alfv\'en speed $V^2_*= B^2_* / 4 \pi \rho_*$, where $B_*$, $V_*$ 
and $\rho_*$ are the poloidal magnetic field, poloidal velocity 
and density along the polar axis at the characteristic radius $r_*$. 
For the magnetic flux function $A$ we define its dimensionless form by
$$
\alpha(R, \theta) = {A(r,\theta)  \over 2 r^2_* B_*}
\,.
$$ 
Note that along the polar axis $\alpha(R, 0) = 0$. To obtain the final
expressions for the physical variables, we make the following crucial
assumptions: 

\begin{itemize}
\item{\it First}, we assume that the Alfv\'en surface is {\it spherical},
$M=M(R)$. Then, according to Eq. (2.1), the density can be expressed as the
product of a function of $R$ [i.e. $1/4\pi M^2(R)$] and a function of $\alpha$
[i.e. $\Psi_A^2(\alpha )$]. Furthermore, we Taylor expand the function
$\Psi^2_A(\alpha)$ to the first order in $\alpha$ such that the  variation of
the density on a spherical surface of given radius $R$ is proportional to the
magnetic flux $\alpha$. 

\item{\it Second}, we assume that the magnetic flux function $\alpha$ is
expressed as the product of a function of $R$ and a function of $\theta$.
Furthermore, for the function of $\theta$ we take a dipolar dependence
with the colatitude $\theta$. This immediately implies that the Alfv\'en cross
sectional area $\pi \varpi_a^2$ of a flux surface is proportional to the
corresponding magnetic flux $\alpha$. Also, the ratio $G^2 = 
\varpi^2 / \varpi_a^2$
of the cross sectional area of the flux tube to the Alfv\'en cross sectional
area of the same flux tube depends solely on the radial distance $R$. 

\item{\it Third}, we assume that the total axial current $I_z$ enclosed by a
flux tube $\alpha$ = const. is proportional to the corresponding magnetic flux.
This assumption fixes the angular momentum integral $L$ (Paper III). Note that at
once the integral of the corotation frequency $\Omega$ follows from its
relation with $L$ at the Alfv\'en distance, $L = \Omega \varpi_a^2$. Note also
that the integrals $L$ and $\Omega$ are chosen such that $L\Psi_A$ and
$L\Omega\Psi_A^2$ contain only first order $\alpha$-terms, in analogy with the
previous assumptions. 

\item{\it Fourth}, we assume that the $\alpha$-dependence of the gas pressure
is similar to that of the density distribution. This means that the pressure is
ultimately a function of the density along a given magnetic surface, a
situation analogous to the often used polytropic assumption. However, this
implicit relationship between pressure and density is much more general than
the somehow artificial polytropic assumption. Contrary to the polytropic
relation, its exact form is not imposed {\it a priori} but is determined by the
full solution. 
\end{itemize}

Altogether, the four main assumptions of this meridionally self-similar 
model can be summarized as follows,
$$
\rho(R,\alpha) = {\rho_* \over M^2(R)} (1 + \delta \alpha)\,,\;
\Psi_A^2=4\pi\rho_*(1 + \delta \alpha)
\,,
\eqno(2.3{\rm a})
$$
$$
\varpi^2 (R, \alpha ) = r^2_*  G^2(R) \alpha\,,\qquad
\varpi_a^2 (\alpha ) = r^2_* \alpha \,, 
\eqno(2.3{\rm b})
$$
$$
L \Psi_A = \lambda r_* B_* \alpha\,,
\qquad
L  \Omega \Psi_A^2  =  \lambda^2 B_*^2 \alpha
\,,
\eqno(2.3{\rm c})
$$
$$
P(R,\alpha) = {1 \over 2} \rho_* V^2_* \Pi(R)(1+ \kappa \alpha)
\,.
\eqno(2.3{\rm d})
$$
The introduced parameters $\delta$, $\kappa$ and $\lambda$ measure the
variation with the colatitude of the density, pressure and rotation,
respectively. A fourth parameter $\nu$ enters from the momentum equation as the
ratio, at the Alfv\'en distance along the polar axis, of the escape speed to the
flow speed there,
$$
\nu^2= {2{\cal G}{\cal M} \over r_* V^2_*}
\,.
\eqno(2.4)
$$ 

\subsection{Magnetic rotator energy}
An important physical quantity in magnetized outflows is the so called {\it
magnetic rotator energy} (Michel 1969, Belcher \& McGregor 1976),
$$
E_{\rm MR} = \Omega L 
\,.
\eqno(2.5{\rm a})
$$
The basal Poynting energy $E_{{\rm Poynt.}, o}$, defined as the ratio of the
Poynting flux density $S_z$ per unit of mass flux density $\rho V_z$, is roughly
equal to the magnetic rotator energy $\Omega L$  \underbar{\it if} at the base
the radius of the jet is much smaller than the Alfv\'en radius ($G_o \ll 1)$
and the Alfv\'en number there is also negligibly small ($M_o \ll 1$), 
$$
E_{{\rm Poynt.},o} ={S_z\over \rho V_z}\Bigl|_{o} = 
\Omega L {1-G_o^2\over 1 -M_o^2} 
\approx \Omega L
\,.
\eqno(2.5{\rm b})
$$

\noindent
Let $E_o$ be the sum of the kinetic, gravitational and thermal energies per
unit mass at the base of the outflow. Then the total available energy for the
outflow at the base 
is $E \approx E_o + \Omega L$. Accordingly, we have an outflow from a
{\it Fast Magnetic Rotator} (FMR) when $E_o \ll \Omega L$ and an outflow from a
{\it Slow Magnetic Rotator} (SMR) in the opposite case of $E_o \gg \Omega L$. 

\subsection{Solving the self-similar MHD equations}

In order to solve the resulting MHD equations, it is useful to introduce 
an extra function $F(R)$ (Papers II and III),
$$
F(R) = 2\left(1- {{\rm d} \, \ln G(R) \over {\rm d} \, \ln R}\right) 
\,.
\eqno(2.6)
$$
\noindent
Evidently, while $G(R)$ defined in Eq. (2.3b) measures the dimensionless
cylindrical radius of a flux tube at the distance $R$, $F(R)$ is simply
giving the expansion factor of the streamlines.  The limiting case $F(R)=0$
corresponds to conical expansion and radial fieldlines, while for $F(R)=2$ we
have cylindrical expansion parallel to the axis (collimation). In between these
two regimes the flow is paraboloidal. 

The above assumptions, Eqs. (2.3), immediately give the components of the velocity and
magnetic fields (Eqs. A.3 in Appendix A). On the other hand, the momentum 
conservation law in
combination with the above assumptions gives four {\it ordinary} differential
equations for the four variables $M^2(R)$, $F(R)$, $\Pi(R)$ and $G(R)$ (see
Appendix A for details).

The complete solution of these equations, from the base of the
outflow to infinity, with the required crossing of all appropriate critical
points, is indeed an interesting undertaking and worth of a separate paper.
Here instead, we shall concentrate on some novel results obtained solely by
solving the equations asymptotically far from the Alfv\'en surface 
($R \gg 1$) {\it and} taking into account the boundary conditions on the 
source. 
  

\section{The energy integral and collimation criterion}

\subsection{The generalized Bernoulli integral}

A nonadiabatic flow of a monoatomic gas with ratio of specific heats $\Gamma
=5/3$ is always heated at a net volumetric rate $q$ 
$$
q= \rho \vec{V} \cdot \left (\nabla h - {{\nabla P} \over {\rho}} 
\right )\,, \quad h = {\Gamma \over \Gamma -1} {P \over \rho}
\,.
\eqno(\hbox {3.1a})
$$
With expressions (2.3a) and (2.3d) for the gas density and pressure, it follows
immediately that this heating can be written as 
(see Sec. 5 in Paper III for details of the derivation), 
$$
{q(R,\alpha)\over \rho(R,\alpha) V_r(R,\alpha)}
= {V_*^2 \over 2r_*} {1 + \kappa\, \alpha
\over 1+\delta\, \alpha} {\cal Q}(R)
\,,
\eqno (\hbox {3.1b})
$$
where the dimensionless specific heating rate 
per unit of radial length along a given streamline is,
$$
{\cal Q}(R) = {1 \over \Gamma - 1 }
\left [ M^2{\hbox {d}\Pi\over \hbox {d}R}
+\Gamma \Pi {\hbox {d}M^2 \over \hbox {d}R} \right]
\,.
\eqno (\hbox {3.1c})
$$
Hence, the generalized classical Bernoulli integral (2.2) takes the 
simpler form
$$
E(\alpha)={1\over 2}V_*^2
     {{\cal E}+ \alpha \Delta {\cal E} \over 1+\delta\alpha }
\,,
\eqno (\hbox{3.2})
$$      
where the two {\it constants} ${\cal E}$ and $\Delta {\cal E}$ represent the
polar specific energy and the variation across a streamline of the specific
energy, respectively (in Paper III ${\cal E}$ was denoted by ${\cal F}_1$, $\Delta
{\cal E}$ by ${\cal F}_2$ and ${\cal Q}$ by ${\cal Q}_1$). 

It is straightforward to show  from Eqs. (2.2), (2.3), (A.3) and
(3.1)--(3.2) that ${\cal E}$ and $\Delta {\cal E}$ have the following
analytical expressions (Paper III) 
$$
{\cal E} = {M^4 \over G^4}
                - {\nu^2 \over R}
                + {\Gamma\over \Gamma - 1} \Pi M^2 
                -\int_{R_o}^{R}{\cal Q}(R)dR
\,,
\eqno (\hbox {3.3a})
$$
$$
\Delta {\cal E}
= {M^4\over R^2 G^2}\left[{F^2\over 4} - 1 \right]
-{\delta\,\nu^2 \over R}
+ {\lambda^2\over G^2}
  \left[ M^2 - G^2 \over 1 -M^2\right]^2
$$
$$
+ 2\lambda^2 \left[1 - G^2\over 1 - M^2 \right]
 +\kappa\left[{\Gamma\over \Gamma - 1} \Pi M^2
        -\int_{R_o}^{R}{\cal Q}(R)dR \right]
\,.
\eqno(\hbox{3.3b})
$$
It is worth to digress for a moment and try to get some insight into the 
physical meaning of these two conserved components of the specific energy, 
 ${\cal E}$ and  $\Delta{\cal E}$.

\subsubsection{Polar specific energy}

In the first expression, Eq. (3.3a), the polar energy flux $\cal E$ is composed of 
four terms which are successively the poloidal (i.e. radial here) kinetic 
and gravitational energies, the enthalpy and the heating along the 
polar axis. 

The polar specific energy ${\cal E}$ can be evaluated at both the base of 
the wind $R_o$ and far from it as $R\longrightarrow \infty$,
$$
{\cal E}  =  - {\nu^2 \over R_o}
             + {\Gamma\over \Gamma - 1} \Pi_o M_o^2 
\,
\eqno (\hbox {3.4a})
$$
$$
= {M_\infty^4 \over G_\infty^4}
                + {\Gamma\over \Gamma - 1} \Pi_\infty M_\infty^2 
                -\int_{R_o}^\infty{\cal Q}(R)dR
\,.
\eqno (\hbox {3.4b})
$$
At the base, wherein the kinetic energy of the outflow is negligible, Eq.
(3.4a) shows that the polar energy has basically two terms: the gravitational
energy and the initial input of thermal energy in the form of enthalpy. On the
other hand, at infinity, Eq. (3.4b), the conserved polar specific energy is
composed of the final kinetic energy along the polar axis and the terminal
enthalpy minus the additional extra heating which the flow has received during
its propagation from $R_o$ to infinity. 

Note that if the wind is cylindrically collimated, $M_\infty$, $G_\infty$ 
and $\Pi_\infty$ have  finite values. In all other cases, 
$M_\infty$ and $G_\infty$ may be unbounded,  although their ratio, which 
is the polar speed in units of the Alfv\'en speed, should remain finite,
$$
{M_\infty^2 \over G_\infty^2} =  {V_\infty \over V_*}
\,.
\eqno (\hbox {3.4c})
$$ 
Moreover the terminal pressure $\Pi_\infty$ vanishes unless the
integral of the heating diverges, a rather unphysical situation corresponding
to an infinite input of heat. 

The conservation of the polar energy simply expresses the fact that the 
flow along the polar axis is {\it thermally 
driven}. Furthermore, from Eqs. (3.4a,b) it 
becomes evident how the conversion of the heat content of the plasma 
into kinetic and gravitational energy maintains the outflow,
$$
{\Gamma\over \Gamma - 1} ( \Pi_o M_o^2 - \Pi_\infty M_\infty^2  )
+ \int_{R_o}^\infty{\cal Q}(R)dR
$$
$$
= {\nu^2 \over R_o} + {M_\infty^4 \over G_\infty^4} 
\,.
\eqno (\hbox {3.5})
$$
In other words, the decrease of the enthalpy 
at infinity together with the external 
heat input integrated along the polar streamline, 
{\it on one hand} lifts the gas out of the gravitational 
potential well and {\it on the other}, gives to it a finite terminal speed. 
Of course, this is nothing more 
than the classical picture of the Parker thermally driven wind.

\subsubsection{Variation of the specific energy across streamlines}

The second conserved component $\Delta {\cal E}$ of the 
specific energy gives the excess or deficit of the volumetric 
total energy $E$ at a nonpolar streamline as compared to the 
corresponding energy at the polar axis and the same spherical 
distance, normalized to the volumetric energy of the magnetic 
rotator. Thus, $\Delta {\cal E}$ has five contributions  
which correspond to the five different terms appearing 
successively in the RHS of Eq. (3.3b). Each one represents 
the variation -- in units of the volumetric energy of the 
magnetic rotator -- between any streamline and the polar axis 
of (i) the poloidal kinetic energy, 
(ii) the volumetric gravitational energy, 
(iii) the azimuthal kinetic energy (which is zero along 
the polar axis), 
(iv) the Poynting flux (which is also zero along the polar 
axis) and 
(v) the thermal content (enthalpy plus heating; see Appendix B for details).

In a more compact way we may write $\Delta {\cal E} $ as follows,
$$
{\Delta{\cal E} \over 2\lambda^2}
= { \rho(R,\alpha)E(\alpha) - \rho(R,{\rm{pole}})E({\rm{pole}})
                \over \rho(R,\alpha)L(\alpha)\Omega(\alpha)}
\,. 
\eqno(\hbox{3.6})
$$
Evidently, ${\Delta{\cal E} / 2\lambda^2}$ represents the variation across the
flow of the total volumetric energy in units of the volumetric energy of the
magnetic rotator. Therefore, the sign of $\Delta {\cal E}$ determines whether
there is a deficit of energy per unit volume (and not per unit mass) along the
polar streamline as compared to the other streamlines (case $\Delta {\cal
E}>0$) or an excess of energy in the polar streamline as compared to the other
nonpolar streamlines (case $\Delta{\cal E}<0$). 

Furthermore, $\Delta {\cal E} / 2\lambda^2$ can be expressed
(see Appendix B for more comments) in terms of the conditions at the
source boundary $R_o$ where the cylindrical radius is $\varpi_o(\alpha)$,
the escape speed $V_{esc, o}$, the polar density $\rho_o({\rm{pole}})$
and the density at any other streamline $\rho_o(\alpha)$: 
 
$$
{\Delta{\cal E}  \over 2\lambda^2} = { \Delta \left[ \rho_o (
E_{{\rm Poynt.}, o} + E_{{\rm R},o} + h_o  + E_{\rm G,o}   )\right] 
\over \Delta(\rho E_{\rm {MR}})_o}
\,,
\eqno (\hbox{3.7a})
$$
where $\Delta(\rho E_{\rm {MR}})_o$ is the variation of the energy of the
magnetic rotator, $\Delta (\rho E_{\rm Poynt.})_o$ is the variation of the
Poynting energy, $\Delta (\rho E_{\rm R})_o$ is the variation of the
rotational energy at the base, $\Delta (\rho E_{\rm G})_o$ is the variation of
the volumetric gravitational energy at the base and $\Delta (\rho h )_o $ is
the variation of the volumetric thermal flux at the base, respectively, 
$$
\Delta (\rho E_{\rm {Poynt}})_o
=\rho_o(\alpha) E_{{\rm Poynt}, o}(\alpha) 
= \rho_o(\alpha) (\Omega L - \Omega^2 \varpi_o^2 )
\,,
\eqno (\hbox{3.7b})
$$
$$
\Delta (\rho E_{\rm R})_o
=\rho_o(\alpha) E_{{\rm R},o}(\alpha) = \rho_o(\alpha){V_{\varphi, o}^2  
\over 2}  = \rho_o(\alpha) {\varpi_o^2\Omega^2  \over 2} 
\,,
\eqno (\hbox{3.7c})
$$
$$
\Delta (\rho E_{\rm G})_o = - {{\cal G}{\cal M} \over r_o}
\left[ \rho_o(\alpha ) - \rho_o({\rm{pole}}) \right]
\,,
\eqno (\hbox{3.7d})
$$
$$
\Delta (\rho h )_o 
=  {\Gamma \over \Gamma - 1}[P_o(\alpha) - P_o({\rm{pole}})]
\,,
\eqno (\hbox{3.7e})
$$
$$
\Delta (\rho E_{\rm MR})_o= \rho_o(\alpha) E_{\rm MR}(\alpha) = \rho_o(\alpha)
L(\alpha) \Omega(\alpha)
\,.
\eqno (\hbox{3.7f})
$$
In this notation, $\Delta$ always denotes a variation across the fieldlines at
a given radial distance $R$, i.e. $\Delta y = y(R,\alpha) - y(R, {\rm pole})$
for  every function $y(R,\alpha)$. 

In Eq. (3.7a) note that (see also Eqs. 2.5) 
$$
E_{{\rm Poynt.}, o} + E_{{\rm R},o}
= E_{\rm {MR}} - E_{{\rm R},o}> 0
\,.
\eqno (\hbox{3.8})
$$
The Poynting flux plus the rotational energy is simply  the energy of the
magnetic rotator minus the rotational energy. This last form is the one used in
Paper III. In other words, and even in the slow magnetic rotator limit, the 
rotational energy never  exceeds the energy of the magnetic rotator. 

\subsection{Energetic definition of Efficient/Inefficient Magnetic Rotators}

At this point we inevitably note that ${\cal E}$ and $\Delta {\cal E}$ are two
inconvenient constants because their absolute values depend on the integration
of the total heating supply and so they can be evaluated only after the problem
has been solved and the required heating can be calculated. However 
these two constants are related to each other. In fact, the last two terms in
the expression of $\Delta {\cal E}$ in Eq. (3.3b), which correspond to the
transverse variations of enthalpy and heating, are identical to the last two
terms of ${\cal E}$ within a factor of $\kappa$. Evidently, this is due to the
assumptions on the pressure and density distribution, Eqs. (2.3a,d). These
initial assumptions imply the existence of an {\it implicit} relationship 
between the latitudinally normalized pressure and density, 
$$
{P(R,\alpha)\over P(R, 0 )}
= 1 + {\kappa\over\delta}
      \left[{\rho(R,\alpha)\over \rho(R, 0)} -1\right]
\,.
\eqno (\hbox{3.9})
$$ 
The situation is akin to the more familiar polytropic 
ansatz, although there the relationship between pressure and 
density is  {\it explicit}.  
In TTS97 the generalized Bernoulli integral has indeed a form 
similar to Eq. (3.2), but the two constants ${\cal E}$ and 
$\Delta {\cal E}$ are not related to each other as in Eqs. 
(3.3), because the spherically symmetric part of the pressure 
is not related to the corresponding nonspherical part.  
For this reason, it was impossible to find a relationship
between $P$ and $\rho$ of the form of Eq. (3.9) and therefore
 any convenient form of the Bernoulli integral.

With this in mind, we can eliminate from the expressions of ${\cal E}$ 
and $\Delta {\cal E}$ in (Eqs. 3.3) the inconvenient enthalpy and heating 
terms (Paper III) by defining the new constant  
$$
\epsilon =\Delta{\cal E}-\kappa{\cal E}
\,.
\eqno (\hbox {3.10})
$$
Now this quantity $\epsilon$, in addition of being a constant for {\it all} 
streamlines,
$$
\epsilon
={M^4\over (GR)^2}\left[ {F^2\over 4} - 1 \right]
- \kappa {M^4\over G^4}
- {(\delta\,- \kappa) \nu^2 \over R}
$$
$$
+ {\lambda^2 \over G^2} \left({M^2-G^2 \over 1-M^2}\right)^2
+ 2\lambda^2{1-G^2 \over 1-M^2}
\,,
\eqno (\hbox {3.11})
$$
 can be calculated {\it a priori} from the conditions at the base of the 
outflow, without a need to know the total input of heating along each line.

A careful look at Eq. (3.11) shows that 
all the transverse variations of the total energy, simply 
reproducing, within a scaling factor $\kappa$, the effect of 
thermally driven winds along the pole (Eq. 3.4), have been removed 
(see Eq. B.7 in Appendix B).

In fact, comparing Eq. (3.11) to Eq.  (3.3b), we see that $\epsilon$ contains 
the same terms as $\Delta {\cal E}$  except the heat content, but with
{\it two} extra terms proportional to $\kappa$. The {\it first} of these two
terms ($\kappa M^4/G^4$) represents simply the transverse variation of the heat
content which is converted into kinetic energy in a thermally driven wind, as
seen by Eq. (3.5). The {\it second} term ($\kappa \nu^2 / R$) is the variation 
with the latitude of the thermal energy which along the pole supports 
the plasma against gravity. 

Let's assume for a moment $\delta = \kappa > 0$, such that the enthalpy and 
the temperature ($\propto P/\rho$) are spherically symmetric.  
Since the pressure is larger on  a nonpolar streamline, we have higher heating 
rate $q$ there: the extra heating converted into kinetic energy is 
$\kappa M^4/G^4$ (Eq. 3.5). In the total
energy variation budget it represents the efficiency of thermal confinement.
Therefore it must be removed from the energy variation in order to
form the constant $\epsilon$. The same holds if $\kappa <0$ except that this
$\kappa$ term will tend to decollimate the  outflow. 

Now, if $\kappa = 0$ and $\delta >0 $ we see that there is an excess of 
gravitational potential $-\delta\nu^2/R$ because the plasma is heavier on a 
nonpolar streamline. 
In order to achieve equilibrium, part of the Poynting flux and part
of the centrifugal
energy must  compensate this term. This reduces the energy available for 
magnetic confinement. If $\delta > \kappa$, we need to correct the previous
argument because part of the weight of the plasma is supported on a non polar 
streamline by an increase of the pressure gradient. This compensation is
exactly $\kappa \nu^2 / R$. Thus the term $-(\delta - \kappa)\nu^2/R$ is the
effective increase of the gravitational potential that must be 
compensated by some non thermal drivers, the magnetic driver for instance.
 It reduces the efficiency of the magnetic rotator to
collimate the flow.
Similar arguments hold if $\delta < 0$ or $\delta < \kappa$.

As in Eqs. (3.7) let's express $\epsilon / 2\lambda^2$  in terms of
the conditions at the source boundary $r_o$ (assuming again that the velocity
is negligible there, see Appendix B for details of the derivation), 
$$
{\epsilon  \over 2\lambda^2} =
{  E_{{\rm {Poynt.}},o} + E_{{\rm R},o} 
+\Delta E_{\rm G}^* 
\over E_{\rm {MR}} }
\,,
\eqno (\hbox{3.12a})
$$
\noindent
where $E_{\rm {Poynt.}}$ and $E_{{\rm R},o}$ have
been already defined. 
$\Delta E_{\rm G}^*$ is the excess or the deficit on a
nonpolar streamline compared to the polar one of the gravitational energy (per 
unit mass) which is not compensated by the thermal driving,
$$
\Delta E_{\rm G}^* 
= - {{\cal G}{\cal M} \over r_o}
\left[  1-{T_o(\alpha)\over T_o({\rm{pole}})} \right]
=  -{{\cal G}{\cal M} \over r_o} {(\delta - \kappa )\alpha 
\over 1 + \delta \alpha } 
\,.
\eqno (\hbox{3.12b})
$$ 
It is indeed the term proportional to $(\delta - \kappa)\nu^2$ in Eq. (3.11)
and the symbol $\Delta$ keeps the same meaning as previously (see Appendix B).

It is worth to remark that this corrected gravitational term plays an important
role in thermally accelerating the flow (Tsinganos \& Vlastou 1988;  Paper I)
because it is proportional to the relative variation of the temperature. We
know from previous numerical studies that $(\kappa - \delta)$ ought to be
negative in order that we have efficient initial acceleration along the polar
axis. This amounts to say that the temperature along the polar axis must be
larger than the temperature along a non polar line. Then, the corrected
gravitational term in Eq. (3.12a) is always negative such that it must be
compensated with part of the initial input of the magnetocentrifugal terms
(Poynting and rotational) at the base of the flow. 

Hence, $\epsilon>0$ means that the magnetocentrifugal terms are dominating the
variation of gravity and that there is some energy left from the magnetic
rotator to collimate the wind. While $\epsilon<0$ means that the magnetic
rotator cannot collimate the wind by itself. Of course the collimating
efficiency of the magnetic rotator may be eventually lowered if there is
further pressure gradient acting outwards in the wind ($\kappa<0$) but
$\epsilon / 2\lambda^2$ really quantifies the original strength of the magnetic
rotator to support the collimation of the flow. 

As a conclusion of this subsection, we may define as 
 {\bf Inefficient Magnetic
Rotators (IMR)}  the magnetic rotators which are not able to confine the flow
through magnetic processes alone and have $ \epsilon < 0$. Conversely
we shall call {\bf Efficient Magnetic Rotators (EMR)} the
 magnetic rotators potentially able to magnetically confine the 
 flow and which have
$\epsilon >0$. We shall further illustrate this definition at the
end of the next subsection. Within this definition the classical {\bf Slow
Magnetic Rotators (SMR)} and {\bf Fast Magnetic Rotators (FMR)} correspond
respectively to (IMR) and (EMR) but only in the limit where all other energies
are distributed in a spherically symmetric manner at the source base.

\subsection{Energetic criterion for cylindrical collimation}

The collimation of an outflow can be either of magnetic, or of thermal origin. 
In the following, we discuss how to measure the distribution 
of the thermal content along and across the flow, before 
reaching some conclusions on the collimation itself.

In a thermally driven wind, all thermal input (internal 
enthalpy plus external heating provided along the flow) is  
not necessarily fully converted into other forms of energy, unless 
the terminal temperature is exactly zero. There always remains 
some asymptotic thermal content in the form of enthalpy. 
Conversely, we can define the heat content that is really 
used by the flow by defining the {\it converted enthalpy}
$$
\tilde h (r,A) = h (r,A)  
+\Theta_r^\infty(A)
-h(\infty,A) 
\,.
\eqno (\hbox{3.13a})
$$
Along a fieldline $A$ and some radius $r$, the converted 
enthalpy $\tilde h$ is simply the sum of 
the enthalpy of the gas at this point and the external heat 
which it will receive on its way to infinity, $\Theta_r^\infty(A)$
(Eq. 2.2b), minus the 
enthalpy that will still remain in the gas asymptotically. 
Note that in the polytropic case this converted enthalpy is 
simply the variation along the flow of the effective enthalpy 
$\bar h_o- \bar h_\infty$, where the adiabatic index $\Gamma$ 
is replaced by some effective $\gamma<\Gamma$, as explained 
in Paper III. Then we can define a constant along each streamline 
$$
\tilde E(A)= {1\over2}V_{\rm p}^2 + {1\over2}V_{\varphi}^2 
- {{\cal G}{\cal M}  \over r} - {\Omega \, \over \Psi_A} \varpi B_\varphi
+ \tilde h  
\,.
\eqno (\hbox{3.13b})
$$
We may also redefine the variation 
across fieldlines of the volumetric energy normalized with 
the energy of the magnetic rotator, but including the 
converted enthalpy which will indeed be used by the flow,
instead of the enthalpy. In other words, we may define a 
new quantity $\epsilon^{\prime}$ in the way we defined $\Delta 
{\cal{E}}$ in Eq. (3.7a), but using the converted enthalpy 
instead of the enthalpy, 
$$
{\epsilon^{\prime} \over 2\lambda^2}
= { \rho(R,\alpha)\tilde E(\alpha) 
- \rho(R,{\rm{pole}})\tilde E({\rm{pole}})
\over \rho(R,\alpha)L(\alpha)\Omega(\alpha)}
\,. 
\eqno(\hbox{3.14a})
$$
Thus we have at the base
$$
{\epsilon^{\prime} \over 2\lambda^2} =
{ \Delta \left[ \rho_o (
E_{{\rm Poynt.},o} + E_{{\rm R},o} 
  +  \tilde h_o   +  E_{\rm G, o} ) \right]   
\over \Delta(\rho E_{\rm {MR}})_o}
\,,
\eqno (\hbox{3.14b})
$$
\noindent
where all the terms have the same meaning as in Eq. (3.7a) 
except the transverse variation of the total converted 
enthalpy of the flow which is simply
$$
\Delta (\rho \tilde h )_o = \rho_o(\alpha)\tilde h_o (\alpha)
                          - \rho_o({\rm{pole}})\tilde h_o({\rm{pole}})
\,.
\eqno (\hbox{3.14c})
$$
Working out this definition together with Eqs. (3.7), (3.3) and (B.5.), we 
find the following relation 
$$
\epsilon^{\prime} =
{\Delta {\cal E}}
-\kappa{\Gamma\over \Gamma - 1} \Pi_\infty M_\infty^2 
+\kappa\int_{R_o}^\infty{\cal Q}(R)dR 
\,.
\eqno (\hbox {3.15})
$$
Thus $\epsilon^{\prime}$ is simply the difference of $\Delta {\cal{E}}$ 
and the total heat content of the flow at infinity. 
As a consequence, in our model $\epsilon^{\prime}$ is a constant 
which can be evaluated {\it a priori} at any $r$ using 
Eq. (3.15). In the particular case of a flow which is 
asymptotically cylindrically collimated ($F_\infty=2$), 
this parameter can be evaluated at infinity in a simple way
(cf. Eqs. 3.3b - 3.15), 
$$
{\epsilon^{\prime} \over 2\lambda^2}=
{{(M^2_\infty - G^2_\infty)^2} \over 2 G^2_\infty(M^2_\infty - 1)^2 }
+ { G^2_\infty - 1 \over M^2_\infty - 1 }
\,.
\eqno(\hbox{3.16})
$$
Note that now in $\epsilon^{\prime}$ there are left only the 
magnetocentrifugal terms: variation of the azimuthal 
kinetic energy and Poynting flux.  The important  result 
is that this new parameter is always positive in a 
cylindrical jet, if the jet is asymptotically superAlfv\'enic, 
$M_\infty>1$, and transalfv\'enic so the asymptotic radius 
is larger than the Alfv\'en radius, $G_\infty>1$. 
In other words, {\sl a necessary condition for achieving 
cylindrical collimation is that $\epsilon^{\prime} > 0$}. 
 
The criterion for cylindrical collimation is thus explicitly 
equivalent to the criterion given in Paper III, except that now we 
have included the thermal contributions: cylindrical 
collimation can be achieved only if there is an excess of 
energy on a non polar line compared to the polar one. 
However, it is not the variation across the lines
of the total thermal energy input that enters in the 
definition of the criterion, 
but the variation of
the thermal energy that is effectively converted into some 
other form of energy between the base and the asymptotics 
(Eq. 3.13). 

Two contributions may arise to give a positive value for 
$\epsilon^{\prime}$: either because the energy of the magnetic rotator 
dominates as in Paper III, or because the thermal contribution 
converted to non thermal energy in the flow is higher outside 
the polar axis.
This last point may be better realized if we note that 
$$
\epsilon^{\prime} \equiv \epsilon + \kappa {V_\infty^2\over V_*^2} 
\,.
\eqno(\hbox{3.17})
$$
Thus, $\epsilon^{\prime}$ splits into two parts. The first is 
$\epsilon$ which is essentially positive when the energy of 
the magnetic rotator dominates (see Paper III and the previous subsection).
The second corresponds to the variation with colatitude 
of the thermal energy that has been converted to kinetic 
energy (see Eqs. 3.5 and 3.4c).

Altogether, there are two ways to have $\epsilon^{\prime} >0$:

\noindent
{\it Either,} when the outflow is magnetically dominated, which 
means that $\epsilon$ is positive and $\kappa$ may be either 
positive (which adds some extra pressure confinement), or 
negative (which corresponds to pressure support of the jet) 
within some limits. 

\noindent
{\it Or}, conversely, when there is a significant contribution 
of the variation of the enthalpy+heating term that is 
converted into kinetic poloidal energy, then 
$\kappa {V_\infty^2/ V_*^2}$ is
large enough which implies $\kappa>0$ while $\epsilon$ may be negative.
This does not necessarily implies that the flow is pressure confined as we 
shall see later.


\section{Asymptotic confinement}  

\begin{figure*}
\epsfig{figure=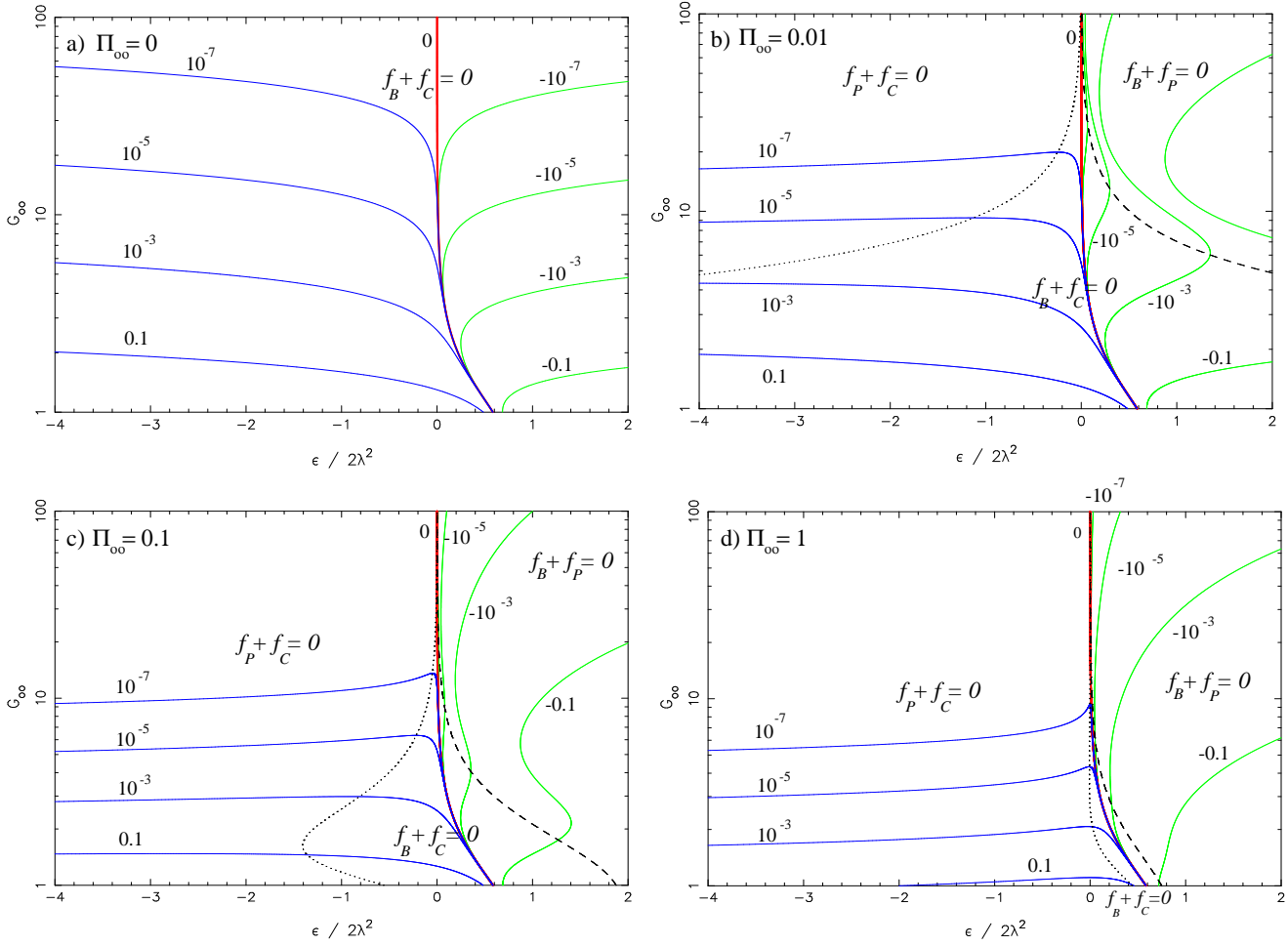}
\caption{
Plots of the asymptotic cylindrical radius normalized to the
cylindrical radius at the Alfv\'en  surface $G_\infty$ 
vs. $\epsilon / 2\lambda^2$ for various values of the final 
pressure: $\Pi_\infty=0$ (panel a),  $=0.01$ (panel b), $=0.1$ (panel c) and
$=1$ (panel d).
Each curve is drawn for a constant value of $\kappa / 2\lambda^2$ 
between $-0.1$ and $0.1$ which labels the curve.
On the left of the dotted line is the domain of pressure confined jets
while on the right of the dashed line is the domain of pressure supported 
jets and in between is the domain of magnetocentrifugal jets.
}
\end{figure*}

\begin{figure*}
\epsfig{figure=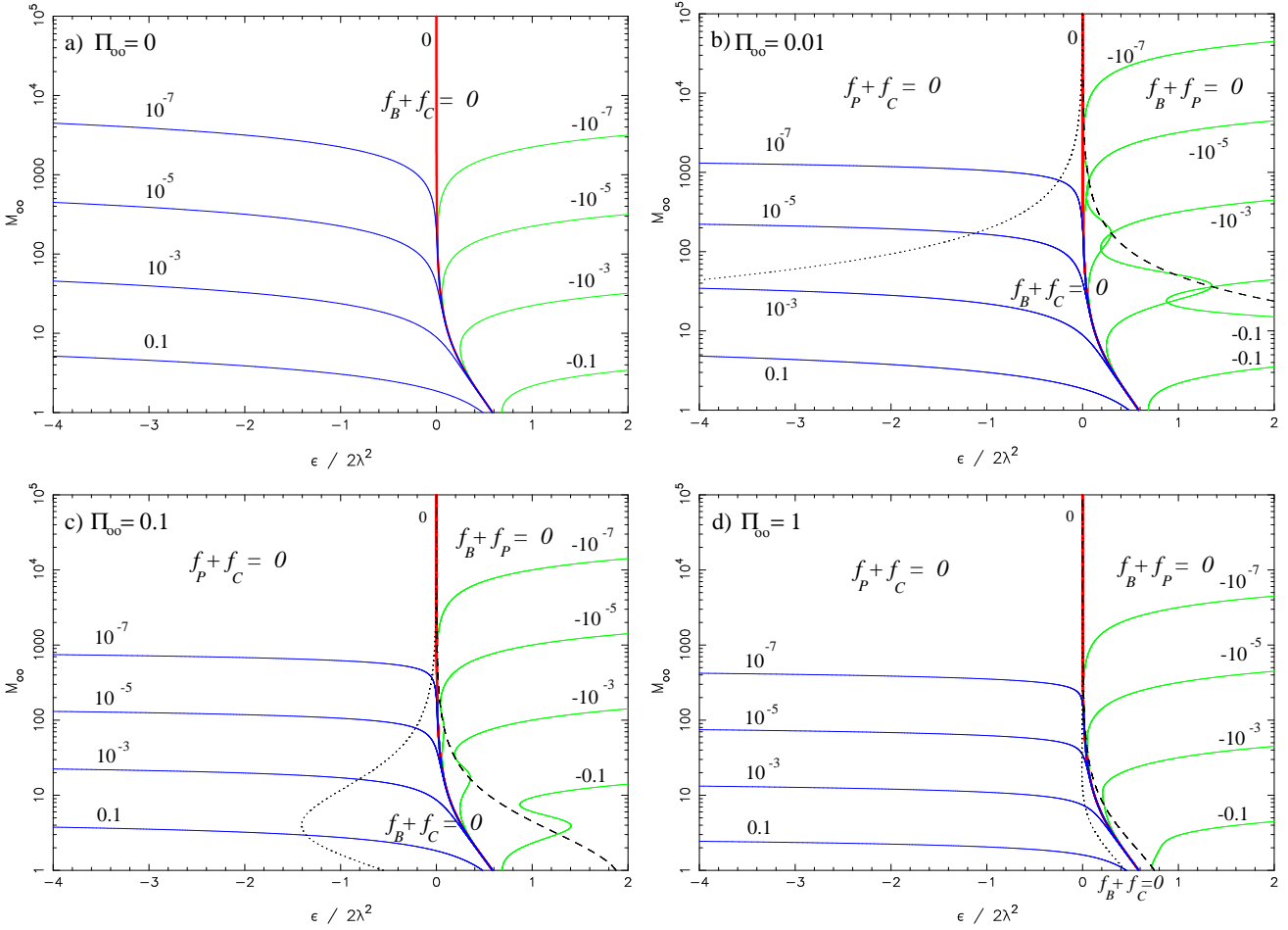}
\caption{
Plots of the asymptotic Alfv\'en number $M_{\infty}$ vs. $\epsilon / 
2\lambda^2$ for the same values of  $\Pi_\infty$ and
$\kappa / 2\lambda^2$ as in Fig. 1.  
On the left of the dotted line 
is the domain of pressure  confined jets
while on the right of the dashed line is the domain of pressure supported 
jets, in between is the domain of magnetocentrifugal jets (except for a small
overlap around the dashed line of panel b).
}
\end{figure*} 

\begin{figure*}
\epsfig{figure=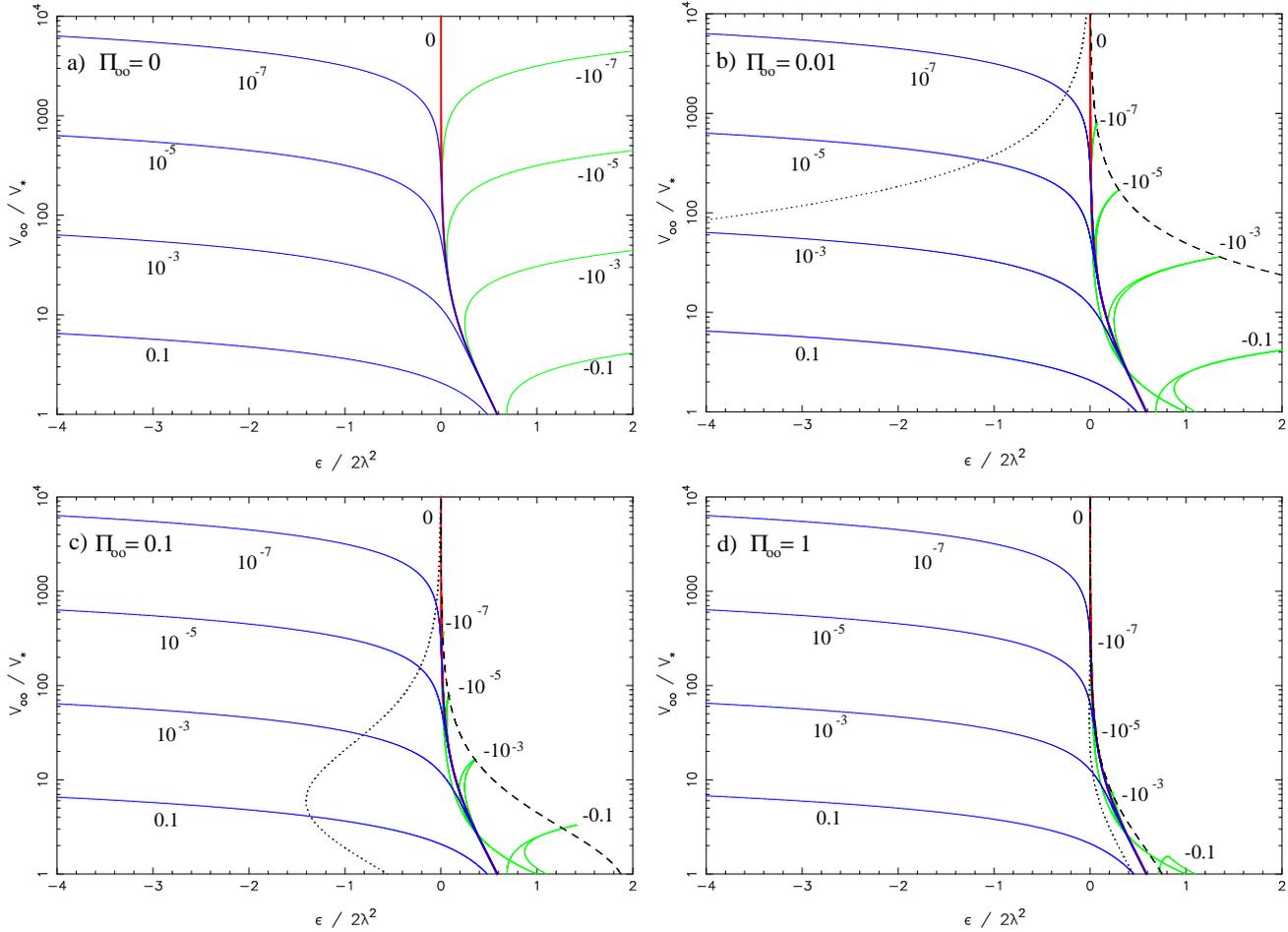}
\caption{
Plots of the asymptotic axial velocity normalized to the 
polar Alfv\'en velocity $V_\infty / V_*$ 
vs. $\epsilon / 2\lambda^2$ for the same values of $\Pi_\infty$ and 
$\kappa / 2\lambda^2$ as in Fig. 1. 
The domain of pressure 
confined jets is clearly on the left side of the dotted line
and the magnetocentrifugal one in between the dotted and dashed line. 
However the pressure supported domain overlaps the magnetocentrifugal one.
}
\end{figure*}

\begin{figure}
\epsfig{figure=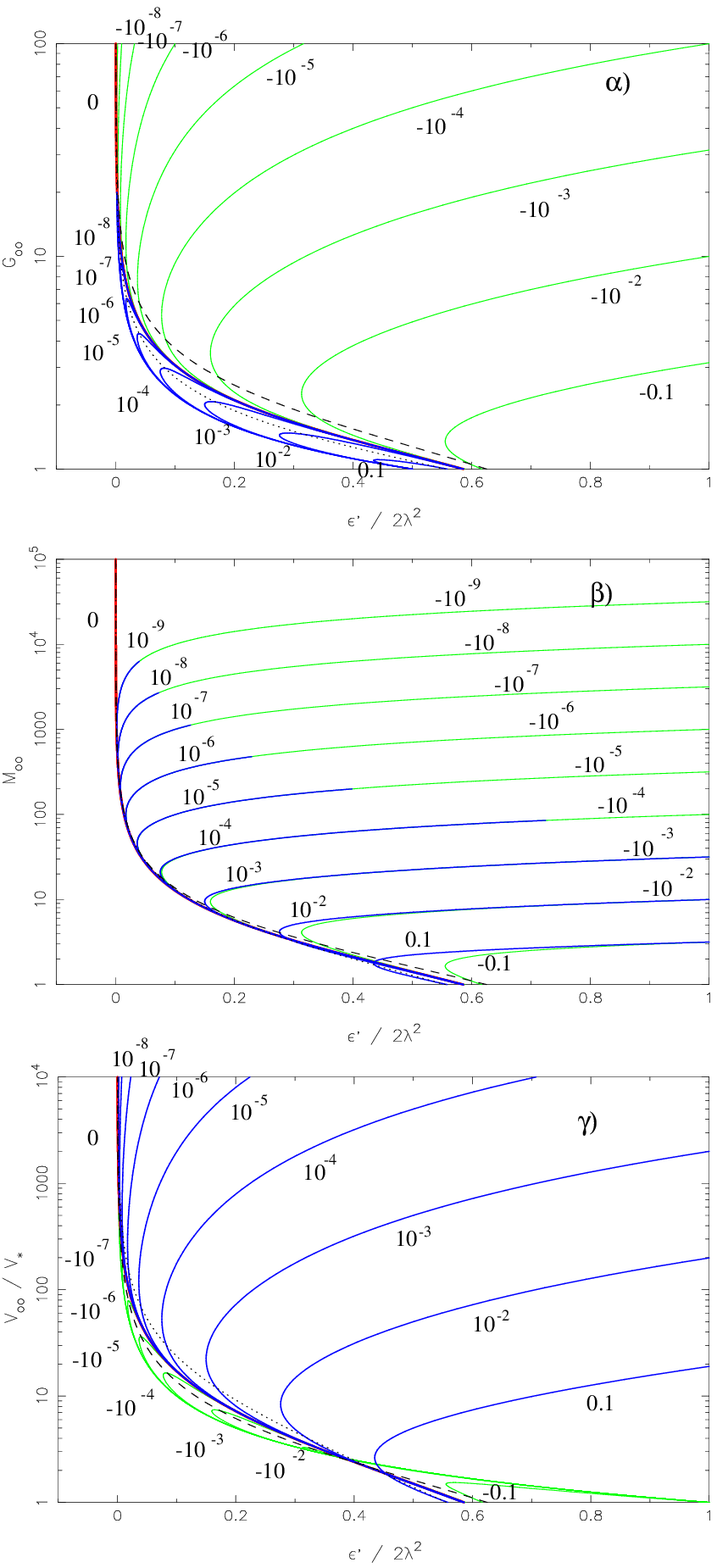}
\caption{Plots of the asymptotic cylindrical radius
(panel $\alpha$), Alfv\'en number (panel $\beta$) and axial velocity (panel
$\gamma$) vs. $\epsilon' / 2\lambda^2$. 
$G_{\infty}$ and $V_{\infty}$ are normalized to the cylindrical
radius and the polar Alfv\'en velocity at the Alfv\'en surface, respectively.
Each curve is drawn for a constant value of $\kappa \Pi_\infty / 2\lambda^2$
(and {\bf not} $\kappa / 2\lambda^2$) between $-0.1$ and $0.1$ which labels the
curve. Curves labeled with $0$ correspond to all curves of panels (a) in
Figs. 1--3, curves labeled with an even 
power of 10 correspond to curves drawn
in panels (c) of Figs. 1--3, while curves labeled with an odd power of 10
correspond to curves drawn in panels (b) and (d) of Figs. 1--3. 
}
\end{figure}

\noindent
We proceed now towards an asymptotic analysis of the equations of motion, in
particular in the case of cylindrical collimation. 

\subsection{Asymptotic equilibrium in cylindrically collimated outflows}

When the equilibrium is asymptotically cylindrical $\epsilon^{\prime} > 0$ as 
discussed above.
Taking the dominant terms in the $\theta$-component of the momentum equation
or,  equivalently by expressing force balance across the cylindrical
fieldlines,  we obtain the condition of MHD equilibrium in the cylindrical 
radius direction ${\hat \varpi}$
expressed by the equation, 
$$ 
\vec{f}_C + \vec{f}_B + \vec{f}_P = 0
\,.
\eqno(4.1)
$$
In the asymptotic regime, the centrifugal ($\vec{f}_C$), magnetic 
($\vec{f}_B$) and gas pressure gradient ($\vec{f}_P$) 
forces have the familiar expressions,
$$ 
\vec{f}_C = {\rho V^2_\varphi \over \varpi}{\hat \varpi}
\,,
\eqno(4.2{\rm a})
$$
$$ 
\vec{f}_B 
= -\left[ { \hbox {d} \over \hbox {d} \varpi} 
             \left( B^2_\varphi \over 8 \pi \right) 
          + {B^2_\varphi \over 4 \pi \varpi} \right] {\hat \varpi}
\,,
\eqno(4.2{\rm b})
$$
$$ 
\vec{f}_P = -{ \hbox {d} P  \over \hbox {d} \varpi } {\hat \varpi}
\,.
\eqno(4.2{\rm c})
$$
In our notation, they can be written as, 
$$ 
\vec{f}_C 
       ={\rho_* V_*^2 \over r_* }  
       { \lambda^2 \over G^3_\infty M^2_\infty} 
       \left({ M^2_\infty - G^2_\infty \over M^2_\infty - 1  }\right)^2 
\sqrt{\alpha} {\hat \varpi}
\,,
\eqno(4.3{\rm a})
$$

$$ 
\vec{f}_B =-{\rho_*V_*^2\over r_*}
         { 2 \lambda^2 \over G^3_\infty } 
           \left( G^2_\infty - 1  \over  M^2_\infty - 1  \right)^2
\sqrt{\alpha}{\hat \varpi}
\,,
\eqno(4.3{\rm b})
$$
$$ 
\vec{f}_P  =-{\rho_*V_*^2\over r_*}
         {\kappa \Pi_\infty\over G_\infty}\sqrt{\alpha} {\hat \varpi}
\,.
\eqno(4.3{\rm c})
$$
Note that always the centrifugal force acts outwards while the total
magnetic force (pinching plus pressure) inwards.  
On the other hand, the last term (not appearing in the  Paper III study) 
is the pressure gradient that acts outwards if the flow is {\it 
overpressured} ($\kappa<0$, i.e. the pressure decreases away from 
the axis). In this case 
the jet is necessarily magnetically confined but either centrifugally 
supported or pressure supported.
Conversely, the pressure gradient acts inwards if  the flow is 
{\it underpressured} ($\kappa>0$, i.e. the pressure increases away 
from the axis). 
In this case the flow is centrifugally supported but may be either 
magnetically confined or pressure confined.

By combining the asymptotic transverse force balance (Eqs. 4.1 - 4.3),
$$ 
-{ 1 \over 2 G^2_\infty M^2_\infty } 
\left( {M^2_\infty - G^2_\infty \over M^2_\infty - 1  }\right)^2 
+
{ 1 \over G^2_\infty } \left( {G^2_\infty - 1  \over M^2_\infty - 1  }\right)^2
$$
$$
+
{\kappa \over 2\lambda^2}{ \Pi_\infty }
=0
\,,
\eqno(4.4{\rm a})
$$
with the expression of $\epsilon$ calculated at infinity (Eq. 3.11)
$$
{\epsilon\over 2\lambda^2} 
= - {\kappa\over 2\lambda^2} {M^4_\infty\over G^4_\infty} 
+ {1 \over {(M^2_\infty - 1)^2}} \times 
$$
$$
    \left[ { (M^2_\infty - G^2_\infty)^2 \over 2 G^2_\infty } 
          + (G^2_\infty - 1)(M^2_\infty -1) \right] 
\,,
\eqno(4.4{\rm b})
$$
we obtain the asymptotic jet radius and  Alfv\'en number as functions of the 
parameters $\epsilon/2\lambda^2 $, $\kappa/2\lambda^2$ and the asymptotic 
pressure $\Pi_\infty $.  Plots of the resulting values of $G_\infty$, 
$M_\infty$ and the axial terminal speed 
$V_\infty/V_*$ {\it vs.} $\epsilon/2\lambda^2$ for four representative values 
of $\Pi_\infty$ (0, 0.01, 0.1 and 1) are shown in Figs. 1--3. In each of 
these panels the values of $G_\infty$, $M_\infty$ and $V_\infty$ are plotted 
for a range of values of the pressure parameter $\kappa /2\lambda^2$ between
$-0.1$ and $0.1$ which label the curves. 

In parallel we also plot (Fig. 4)  $G_\infty$, $M_\infty$ and  $V_\infty/V_*$ 
{\it vs.} $\epsilon^{\prime}/2\lambda^2$ using Eq. (3.17) to determine  
$\epsilon^{\prime}/2\lambda^2$. Note that (see Eqs. 4.4) $G_\infty$, 
$M_\infty$ and  $V_\infty/V_*$  depend only on $\epsilon'/2\lambda^2$ and  
$\kappa \Pi_{\infty} /2\lambda^2$. So conversely to Figs. 1--3 where each 
curve is drawn
for given values of $\Pi_\infty$ and  $\kappa /2\lambda^2$ independently, 
the curves of Fig. 4 are drawn for a constant and unique value of 
$\kappa \Pi_\infty /2\lambda^2$ which labels
 the curve. 
In Fig. 5 we make an explicit comparison of the plot  $G_\infty$ vs.  
$\epsilon/2\lambda^2 $ of Fig. 1a and  the corresponding curve 
$G_\infty (\epsilon'/2\lambda^2) $ for $\kappa \Pi_{\infty} /2\lambda^2 = 0 $.

In these plots we may find three different 
regimes for the asymptotic state of the collimated outflow according to the 
various confinement and support conditions across the jet.

\subsection{ Magnetocentrifugal jets with $f_C= |f_B|$}

This is the case when the pressure gradient is exactly zero 
$f_P= 0$
i.e., the pinching magnetic force is balanced by the inertial 
(centrifugal) force alone. The two following cases correspond to this 
situation.

\subsubsection{Spherically symmetric pressure, $\kappa = 0$} 

The pressure is everywhere spherically 
symmetric. The terminal value of the pressure does 
not affect the asymptotic equilibrium in the jet, regardless if it is finite 
($\Pi_\infty\neq 0$) or zero ($\Pi_\infty=0$). This situation corresponds to 
the thick solid curve labeled 0 in Figs. 1--3. 
This special and simplest case  has been already discussed in detail 
in  Paper III 
where it was found that the outflow collimates into a cylindrical jet only for
 $\epsilon >0$, because in this case $\epsilon=\epsilon^{\prime}$. 
For a given $\lambda$, if the flow remains superAlfv\'enic, $M_\infty > 1$,
an upper limit exists for $\epsilon$. For $\epsilon/2\lambda^2 \longrightarrow 
\epsilon_{\rm max}/2\lambda^2 = (2-\sqrt{2}) \approx 0.586 $, then
$M_\infty \longrightarrow 1$ and $G^2_\infty \longrightarrow 1$ (see 
Figs. 1 and 2). 
As $\epsilon$ decreases from $\epsilon_{\rm max}$ the jet's radius, 
Alfv\'en number and terminal speed increase. Finally, 
as $\epsilon \longrightarrow 0$, $G_\infty\longrightarrow \infty$,
$M_\infty \longrightarrow \infty$ and $V_\infty\longrightarrow \infty$. 
The streamlines become conical and the asymptotic speed diverges.
No collimated solutions at all exist for $\epsilon < 0$, as is evident 
from the plots.

\subsubsection{Vanishing asymptotic pressure, $\Pi_\infty=0$} 
The asymptotic gas pressure is zero such
that we have again $f_P =0$, even though $\kappa \neq 0$. This is the
case shown in Figs. 1a--3a where, besides the collimated solutions of the
 Paper III case obtained for $\epsilon >0$ and $\kappa = 0$ 
 (the thick solid branch), 
we have now collimated solutions for practically all values of $\epsilon$,
$-\infty < \epsilon < + \infty$. More specifically, collimated solutions are
found to the left of the thick solid branch for $\kappa >0$ (mainly for 
$\epsilon < 0$) and to the right of the thick solid branch for $\kappa <0$. 

All branches converge for positive $\epsilon$ and small $G_\infty$ towards the
thick solid line. Physically this corresponds to the limit (see Eqs. 4.4b and 
3.17)
where $\kappa V^2_\infty/ V^2_*$ becomes negligible and the magnetocentrifugal
collimation arises because of the central EMR as in  Paper III
($\epsilon^{\prime}\approx\epsilon$). Thus we can compare Figs. 1a--3a to 
the related curve $\kappa \Pi_\infty / 2\lambda^2 = 0 $ of Figs. 4 and 5b.
As $\Pi_\infty=0$, $G_\infty$, $M_\infty$ and $V_\infty/V_*$ depend only on 
$\epsilon^{\prime}$, namely
$$
M_\infty=\sqrt{2}\left[{2\lambda^2\over\epsilon^{\prime}}-1\right]\,,\quad
G^2_\infty= {4\lambda^2\over\epsilon^{\prime}}
\left[{2\lambda^2/\epsilon^{\prime}}-1\over 
{4\lambda^2 /\epsilon^{\prime}}-1\right]
\,.
\eqno(4.5)
$$
This is identical to Eqs. (5.13) and (5.14) in  Paper III with 
$\epsilon^{\prime}$ 
replacing $\epsilon$.

However, even with $\epsilon <0$, we have $\epsilon^{\prime} >0$ provided that 
$\kappa V^2_\infty/ V^2_*$ is larger than $|\epsilon|$. In fact with 
$M_\infty > G_\infty$ ($\Leftrightarrow V_\infty > V_*$), the ratio 
${V^2_\infty/ V^2_*}$ is 
large and positive and can compensate all negative values of $\epsilon$ 
even for small values of $\kappa$. In this way, even as $\epsilon 
\longrightarrow -\infty$, $\epsilon^{\prime} >0$ and collimated solutions are 
obtained, albeit with rather large radii $G_\infty$. In physical terms, this 
corresponds to a situation where the central source is an IMR and cannot 
collimate the flow through magnetic processes alone. Nevertheless, {\it first} 
it is possible that the
conversion of thermal energy into kinetic energy is very efficient 
${M^4_\infty/ G^4_\infty} \gg 1$. {\it Second}, it can be more efficient on a 
non polar streamline than on the polar one  if there is more thermal energy in 
the nonpolar streamlines than in the polar one ($\kappa>0$). In this case then 
as the flow expands it will build up pressure gradients that will force the lines 
to bent towards the axis. Once the thermal energy
is converted into kinetic energy and the pressure is becoming negligible 
the magnetocentrifugal forces will dominate. However the collimation is
obviously less efficient (larger $G_\infty$) than that produced by a central 
EMR (Fig. 1a). Therefore, this excess of thermal 
energy induces the collimated character of the solution through the 
energy integral, even though the corresponding pressure force 
$f_P$ does not enter directly into the asymptotic force balance condition. 
The term ``magnetocentrifugal confinement'' can be used for all 
$(\epsilon / 2 \lambda^2 , \kappa/ 2 \lambda^2)$ values of Figs. 1a--3a. 
 
For $\kappa <0$, in order to keep $\epsilon^{\prime} >0$, larger values of 
$\epsilon$ are required, in comparison to the simple $\kappa =0$ case. In 
physical terms this is so because now there is a deficit of thermal energy 
along the nonpolar streamlines in comparison to the polar one and therefore 
the star has to be a more efficient magnetic rotator (larger values 
of $\epsilon >0$) in order to have a collimated outflow. This trend is 
shown by the grey branches to the right of the thick solid branch 
corresponding to $\kappa = 0$ (Figs. 1a--3a). 

\begin{figure*}
\epsfig{figure=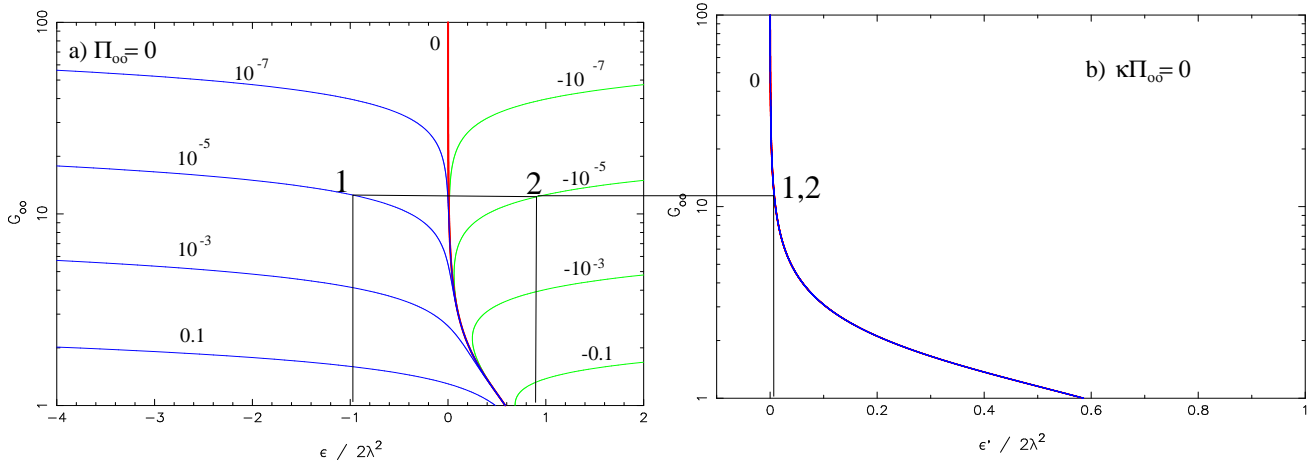}
\caption{
Comparison of the plots of the asymptotic cylindrical radius 
normalized to the cylindrical radius at the Alfv\'en  surface $G_\infty$ vs. 
$\epsilon / 2\lambda^2$ (panel a) and vs. $\epsilon' / 2\lambda^2$ (panel b).
Each curves on the left panel (a) are drawn for a constant value of 
$\kappa / 2\lambda^2$ which labels the curve. All these curves correspond to a
unique curve on the right panel (b) because they all have the same value 
$\kappa \Pi_\infty / 2\lambda^2 = 0 $. We see that for 
each value of $\epsilon^{\prime} /2 \lambda^2$, we 
can find the same value for $G_\infty$ for a pair of values of 
$\epsilon / 2 \lambda^2 $ and 
$\kappa/ 2 \lambda^2$ which have approximately the same magnitude but opposite sign (points
 1 and 2).}
\end{figure*}

Note that for $\kappa>0$, for each value of $\epsilon/ 2 \lambda^2 $ 
there exists a single
value of $G_\infty$, $M_\infty$ and  $V_\infty$. On the other hand for 
$\kappa <0$, more values of $G_\infty$, $M_\infty$ and  $V_\infty$ correspond 
to the same 
value of $\epsilon$. However the lower value of $G_\infty$ is practically 
coincident with that of $\kappa=0$: in other words, thermal decollimation is 
negligible, similarly to the case $\kappa=0$ and 
 $\epsilon^{\prime}\approx\epsilon$. 
The upper values of the branches correspond to efficient conversion of 
thermal energy into polar acceleration but with a deficit along non polar lines.
It induces a more drastic conversion of the Poynting flux  into acceleration 
along non polar lines thus reducing the efficiency of the 
collimation (Fig. 1, $G_\infty$ larger) despite the fact that the central
object is an EMR ($\epsilon >0$). 

This is illustrated in Fig. 5 where the asymptotic 
cylindrical radius $G_\infty$ is plotted versus 
$\epsilon/2\lambda^2$ and $\epsilon^{\prime}/2\lambda^2$ for 
various $\kappa/2\lambda^2$ and $\Pi_\infty = 0$. The higher value of
$G_\infty$ corresponds to a higher conversion of thermal energy that 
decollimates the wind which is balanced by a strong EMR as we explained. For 
some value of $\epsilon^{\prime}=\epsilon+\kappa{V_\infty^2/V^2_*}$, we can 
find 
the same value for $G_\infty$ for a pair of values of $\epsilon$ and $\kappa$ 
which have approximately the same magnitude but opposite sign (see Fig. 5, 
points 1 and 2). 

Note that a situation with $\Pi_\infty=0$ corresponds to
very specific initial conditions of integration that may not be easily 
fulfilled for a cylindrically collimated flow. In particular $\Pi_\infty=0$
would imply $T_\infty=0$ and the existence of some efficient cooling.
 
\subsection{Magnetocentrifugal jets with $f_C \approx |f_B|$}

In addition to the case wherein $\Pi_\infty=0$ shown in Figs. 1a--3a,
where we have an {\it exact} magnetocentrifugal equilibrium everywhere, 
approximate magnetocentrifugal equilibrium conditions also exist for 
$\Pi_\infty\neq 0$. This is shown in Figs. 1b--d, 2b--d, 3b--d and 4, 
in a region
adjacent to the thick solid curve obtained for $\kappa=0$, between the dashed 
and the dotted lines.
 
\subsubsection{Underpressured jets, $\kappa>0$}

Following a branch of $\kappa>0$ on the left side of the limiting curve 
$\kappa=0$, we see from the plots of the various forces shown in Fig. 6, 
that as $\epsilon / 2 \lambda^2$ decreases the magnetic confinement 
of the jet is  replaced by a pressure confinement, as expected. 
This transition from magnetic confinement to pressure confinement can be 
found by writing $f_P=f_B$, or equivalently, $|f_B|=f_C/2$ which gives
$$
G_\infty^2={M_\infty^2 + 2M_\infty \over 1+ 2 M_\infty}
\,.
\eqno(4.6)
$$ 
By inserting  this relation in Eq. (4.4b), we obtain the dotted line of 
Figs. 1--4.
 Thus, for some finite value of $\Pi_\infty$, for each 
positive value of $\kappa/2\lambda^2$, there exists a single value of 
$\epsilon/2\lambda^2 = \epsilon_{P-B}/2\lambda^2$ 
located on the dotted line where there is an equal contribution by the 
magnetic and gas pressure forces in confining the jet against the outwards 
inertial (centrifugal) force. On the left side of the dotted line   
the jet enters the regime of gas pressure confinement, which we shall 
further discuss in the next subsection.  

We note that the limit between the two confinement regimes is close to 
the maximum of the cylindrical radius $G_\infty$ as a function of 
$\epsilon/ 2\lambda^2$, for each value of $\kappa/ 2 \lambda^2$ 
(Figs. 1b--1d). 
The limit is
also very close to the minimum value that $\epsilon^{\prime} / 2 \lambda^2$ 
can achieve for
 a given value of $\kappa/ 2 \lambda^2$ (Fig. 4$\alpha$, thick solid lines). 
On the other hand, $M_\infty$ and $V_\infty$ are monotonic functions of 
$\epsilon/2\lambda^2$. 

This maximum radius of the jet can be calculated formally from Eq. (4.4b),  
$$
G_\infty^2=M_\infty^2 \, {2 M_\infty^2
+ \sqrt{4M_\infty^6 -11M_\infty^4 +10M_\infty^2 - 3}
\over 4M_\infty^4-3M_\infty^2+1}
\,,
\eqno(4.7)
$$ 
which can be combined with Eq. (4.4a) to give the radius as a function of 
the parameters $\epsilon/2\lambda^2$, $\kappa/2\lambda^2$ and $\Pi_{\infty}$. 

It is interesting to note that, in the limit of large Alfv\'en numbers 
($M_\infty\gg 1$) the two values of $G_\infty$ given by Eqs. (4.6) and 
(4.7) coincide and to first order we have
$$
G_\infty^2\approx {M_\infty\over 2} \approx \left(\lambda^2\over 8\kappa 
\Pi_\infty\right)^{1/3}
\,.
\eqno(4.8{\rm a})
$$
Then the asymptotic velocity along the polar axis is (see Eq. 3.4c)
$$
{V_\infty\over V_*}\approx \left(8\lambda^2\over \kappa 
\Pi_\infty\right)^{1/3}\,, \quad \lambda = {\Omega ({\rm pole}) 
r_*\over V_*}
\,.
\eqno(4.8{\rm b})
$$
The terminal speed has the same dependence on the dimensionless 
rotational speed $\lambda$ 
with Michel's minimum energy solution for cold 
magnetic rotators wherein 
$V_\infty /V_*= \lambda^{2/3}$. 
In the present case however, the asymptotic speed is enhanced by the 
factor $[8/\kappa\Pi_{\infty}]^{1/3}$. For small 
values of $\kappa$ and $\Pi_{\infty}$ of the order of 1, this is indeed a 
rather large enhancement. This increased terminal speed simply reflects the 
transformation to asymptotic kinetic energy of the enthalpy and added 
thermal energies. 
Note also that when $\Pi_{\infty} \longrightarrow 0$, $V_{\infty} 
\longrightarrow \infty$. This is expected and the situation is similar to 
the radial outflow studied in Tsinganos \& Trussoni (1991) where the 
terminal speed is   
$$
{V_\infty\over V_*}\approx 6\left( \lambda^2 \ln R\right)^{1/3} 
\,.
\eqno(4.9)
$$

\subsubsection{Overpressured jets, $\kappa<0$}

The regime of magnetocentrifugal equilibrium extends also to the right of 
the thick solid line $\kappa =0$ and up to the dashed line in Figs. 1 and 2. 
Beyond this line the jet enters in the regime of gas pressure support 
which we shall discuss later.
\begin{figure*}
\epsfig{figure=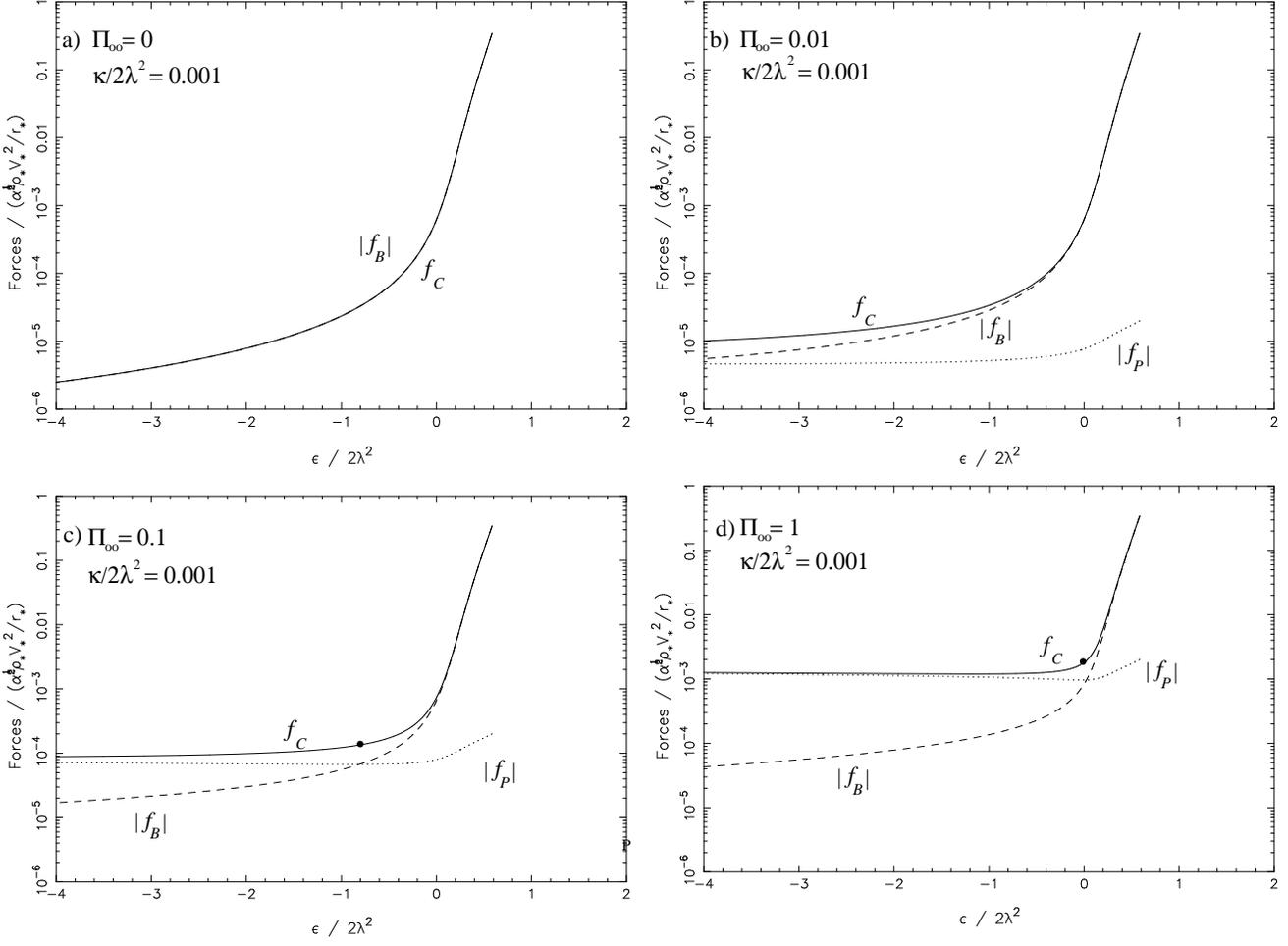}
\caption{Asymptotic transversal forces vs. $\epsilon/2\lambda^2$ 
for a positive value of $\kappa$, $\kappa/2\lambda^2=0.001$: centrifugal
force ($f_C$, solid line), magnetic force ($f_B$, dashed line) and pressure 
gradient ($f_P$, dotted line). The asymptotic pressure is $\Pi_\infty =0$ 
(panel a), $=0.01$ (panel b), $=0.1$ (panel c) and $=1$ (panel d).
The bullet marks the value of the 
centrifugal force $f_C$ for which $f_P=f_B$. 
}
\end{figure*}
\begin{figure*}
\epsfig{figure=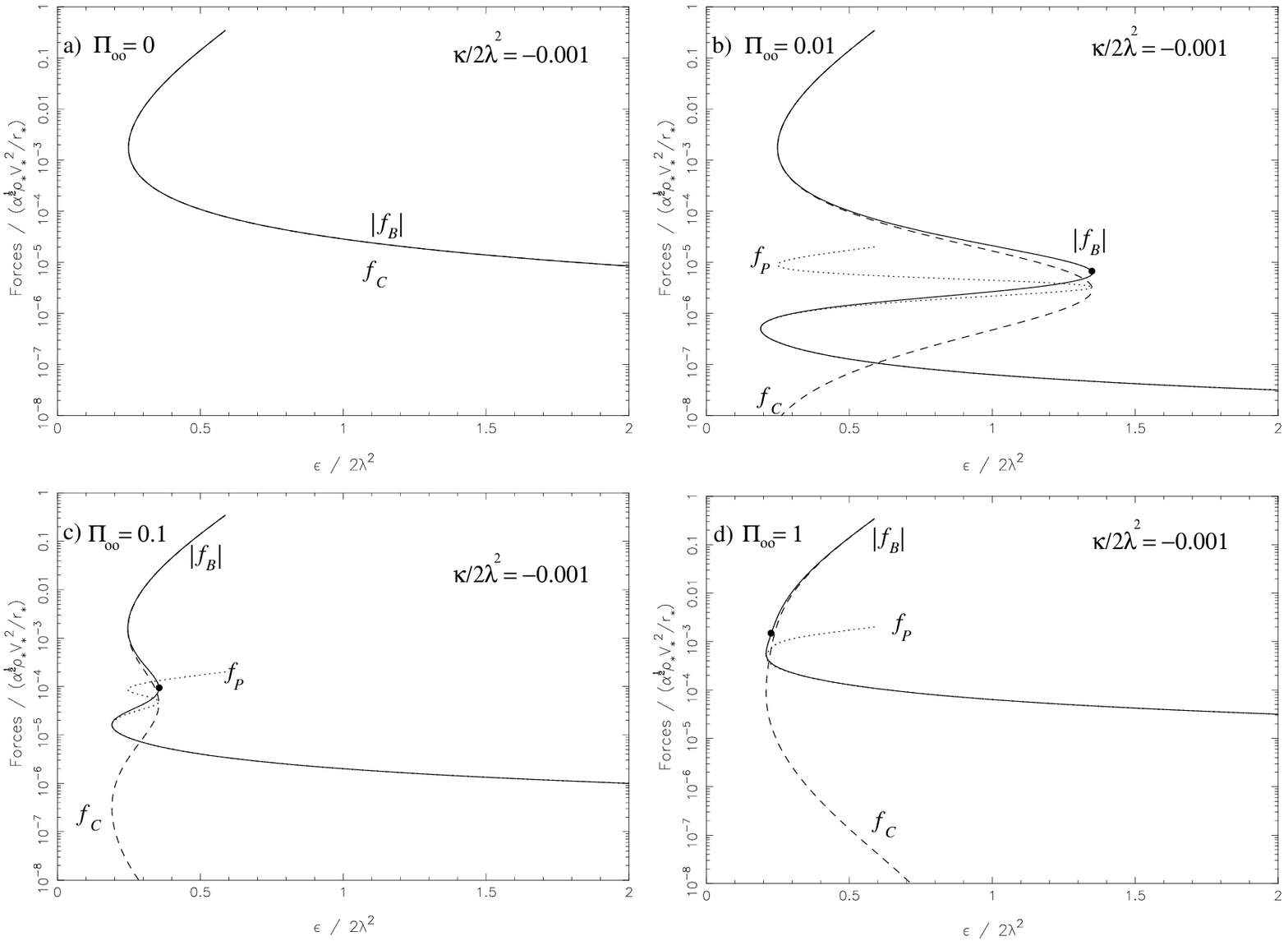}
\caption{
The same as Fig. 6, for a negative value of $\kappa$,  
$\kappa/2\lambda^2=-0.001$ but the magnetic
force ($f_B$) corresponds to the solid line, the centrifugal force ($f_C$) to
the dashed one  and the pressure gradient ($f_P$) to the dotted one.  
The bullet marks the value of the  magnetic force $f_B$ for which $f_P=f_C$. 
}
\end{figure*}
Following a branch of $\kappa<0$ on the right side of the limiting curve 
$\kappa=0$, Fig. 7 
illustrates how the centrifugal force decreases and is 
progressively dominated by the pressure gradient. The transition from 
centrifugal support to pressure support can be estimated by writing 
$f_P=f_C$, or equivalently, $|f_C|=f_B/2$ which gives
$$
G_\infty^2=M_\infty
\,.
\eqno(4.10)
$$ 
This relation can be combined with Eq. (4.4b) to give the dashed 
limiting line (Figs. 1--4) where 
$\epsilon/2\lambda^2 = \epsilon_{P-C}/2\lambda^2$. 
For $G_{\infty} \gg 1$ this limit coincides also approximately with the 
maximum of the velocity 
on an assumed $\kappa/2\lambda^2$ branch (Fig. 3), given by

$$
{V_\infty \over V_*} \biggr |_{\rm max} = {1 \over {2 G^2_{\infty}}} 
\times \left [ G_\infty^4 - 4 G_\infty^2 + 5 \right.
$$
$$ 
\left. + 
\sqrt{G_\infty^8 - 8 G_\infty^6 + 26 G_\infty^4 - 40 G_\infty^2 + 21 } \right ] 
\,,
\eqno(4.11)
$$  
(in this domain the curves of the jet radius $G_\infty$ and Alfv\'en number
$M_\infty$ are monotonic with $\epsilon/2\lambda^2$). We remark that the curve
of maximum velocity (\~ dashed line) also corresponds to the minimum value 
of $\epsilon^{\prime}
/ 2\lambda^2$ for a given value of $\kappa/2\lambda^2$, in the limit of large
Alfv\'en number (see Fig. 4$\gamma$, thin solid lines). 

For $G_{\infty} \gg 1$ we have from Eqs. (4.10):
$$
{ V_\infty \over V_*} \biggr |_{\rm Max} \approx \left ( {{2 \lambda^2} \over
{|\kappa| \Pi_{\infty}}} \right )^{1/3}
\,,
\eqno(4.12)
$$  
i.e. the same scaling law with $|\kappa| /2
\lambda^2$ and $\Pi_{\infty}$ holds for the maximum velocity as for the 
asymptotic velocity at maximum radius for $\kappa >0$ (Eq. 4.8b).             

\subsubsection{$\epsilon_{P-B}<\epsilon<\epsilon_{P-C}$}

In the intermediate region bounded by the two curves $\epsilon_{P-B}$ and
$\epsilon_{P-C}$, the jet is magnetocentrifugal. Solutions very close to the
thick solid line correspond to an efficient collimation by the EMR with negligible
thermal contributions, as we already discussed. For small values of
$\Pi_\infty$ (e.g., Figs. 1b--3b) this area is surrounded by solutions with
important thermal energy conversion but small asymptotic pressure gradients
similar to the extended branches of Figs. 1a--3a. 
 However, for larger  $\Pi_{\infty}$ the jets can be in 
magnetocentrifugal regime only for a narrow range of values of 
$\epsilon /2 \lambda^2$, around  the line $\kappa=0$ (see Figs. 1c--3c and 
1d--3d: the region between the 
dotted and dashed  lines shrinks by increasing $\Pi_{\infty}$).

\subsection{Pressure confined jets $f_C \approx |f_P|$ }
 
The more negative $\epsilon/2\lambda^2$ becomes, the weaker are the magnetic
pinching forces (the less efficient is the magnetic rotator). Thus, for
$\epsilon < \epsilon_{P-B}$ (left of the dotted line on Figs. 1b--1d, 2b--2d,
3b-3d and 4$\alpha$) 
the magnetic pinching force has dropped to very small
values in comparison to the gas pressure force $f_P$, such that now $f_P$ alone
confines the jet against the inertial force (Fig. 6). The situation is similar
to the case of the prescribed streamlines studied in TTS97 ($q>0$, in the
 upper branch of curves in the right panel of Fig. 1 in this paper).  

The asymptotic radius of the jet and its Alfv\'en number are sensitive to 
the nature of the asymptotic confinement. They strongly depend on the value 
of $\Pi_\infty$ (e.g., Figs. 1--2). At the same time, the terminal speed 
$V_\infty/V_*$ is almost independent of the value of the terminal gas pressure 
$\Pi_\infty$ (e.g., Fig. 3) for strongly negative values of  
$\epsilon/2\lambda^2$. This can be understood from Eq. (4.4b) where we see 
that in the limit of negative $\epsilon/2\lambda^2$ the first term of the 
right hand side dominates and thus the square of the terminal speed is roughly
 given by  the ratio $|\epsilon|/\kappa$.

For $\epsilon\ll\epsilon_{P-B}$ we are basically entering the hydrodynamic  
regime studied in Paper I. In the limit of $\epsilon\rightarrow -\infty$ and 
finite asymptotic pressure, the jet is strongly pressure confined such that 
$G_\infty(\epsilon\rightarrow -\infty)<1$ 
(cf. Figs. 1b--1d, 2b--2d, 3b--3d), i.e., the 
solution becomes unphysical. As in   Paper I we find that the most physically 
interesting hydrodynamic solution is obtained for vanishing terminal 
pressure with conical asymptotics wherein 
$G_\infty(\epsilon\rightarrow -\infty)\rightarrow \infty$ (e.g., Fig. 1a 
-- 3a).

We may see the hydrodynamical limit as the most extreme one. Nevertheless,
even with a non vanishing magnetic field, we note that the more efficient is
the pressure confinement of the jet (the more negative is $\epsilon/2\lambda^2$)
the larger is the gap between the value of the jet radius $G_\infty$ obtained 
for $\Pi_\infty=0$ and the one obtained for $\Pi_\infty\gapp 0$ 
(Figs. 1a and 1b), for a given value of $\kappa/ 2 \lambda^2$. 
The difference is even larger when $\kappa/ 2 \lambda^2$ takes small values.
In such a case the magnetocentrifugal forces are very weak (Fig. 6) 
and the equilibrium is  very sensitive to small changes in the pressure 
gradient. 

\subsection{Pressure Supported Jets with $f_P \approx |f_B|$}

This last case occurs when the centrifugal forces are negligible
i.e., the jet is confined by magnetic forces and is supported by the gas 
pressure gradient (for $\epsilon > \epsilon_{P-C}$, right side of 
dashed line, but only in Figs. 1b--d, 2b--d and 4$\alpha$). 

Now the inertial force has dropped to very small values in comparison to the
gas pressure gradient force  such that $f_P$ alone supports the jet
against the magnetic pinching force (see Fig. 7). 
In this domain, for a given
value of $\kappa/2\lambda^2$, the centrifugal force exactly vanishes for
$M_\infty = G_\infty$ (Eq. 4.3a), when
$V_\infty=V_*$. It simply
states that a jet with zero asymptotic centrifugal force has no net
acceleration between the Alfv\'en surface and infinity. At this particular
point $\epsilon/2\lambda^2 =(\epsilon/2\lambda^2)_0$ where the asymptotic
centrifugal force is exactly zero and $V_\infty=V_\star$, we obtain from Eq.
(4.4b), 
$$
\left(\epsilon \over 2\lambda^2\right)_0 +{\kappa\over 2\lambda^2} = 1
\,. 
\eqno (4.13)
$$
Since $\kappa/2\lambda^2$ is usually rather small, it follows that 
$(\epsilon/2\lambda^2)_0 \approx 1$. There, we can say that the jet is 
exactly supported by the pressure gas gradient alone. If we start at 
$(\epsilon/2\lambda^2)_0$ and move toward larger values of 
$\epsilon/2\lambda^2$, along the branch $\kappa/2\lambda^2 =const < 0$, 
$M_\infty$ and $G_\infty$ increase, rotation changes sign (See Eq. A.3f)
and the asymptotic velocity is less than the velocity at the Alfv\'en 
point (Fig. 3) which means that the outflow is decelerated though it remains
superAlfv\'enic (we do not consider in the present analysis ``breeze'' 
solutions that are always subAlfv\'enic). 
 
By starting at $(\epsilon/2\lambda^2)_0$ and moving in the opposite 
direction toward smaller values of $M_\infty$ and $G_\infty$ along the 
branch $\kappa/2\lambda^2 =const < 0$, the inertial force 
remains negligible in comparison to the gas pressure gradient, up to the
dashed line wherein we enter the magnetocentrifugal domain.  


\section{Oscillations in the jet's width} 

\begin{figure*}
\epsfig{figure=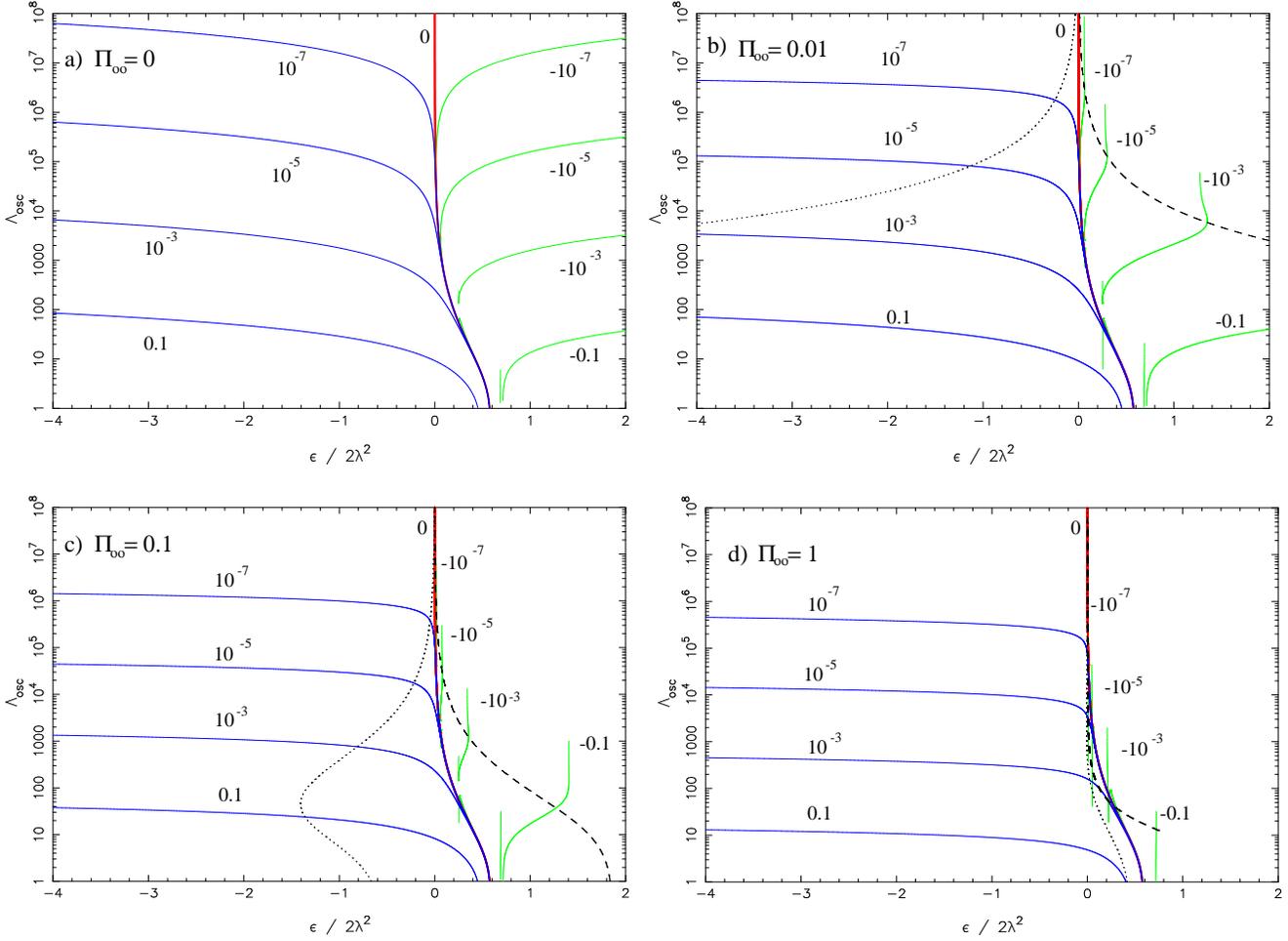}
\caption{Plots of the wavelength of the oscillations $\Lambda_{\rm osc.}$ 
in units of $r_*/\lambda$ vs. $\epsilon/2\lambda^2$  
for the same values of $\Pi_\infty$ and $\kappa/2\lambda^2$ as in Figs. 1--3. 
On the left of the dotted line
is the domain of pressure confined jets
while on the right of the dashed line is the domain of pressure supported 
jets and in between is the domain of magnetocentrifugal jets.
}
\end{figure*}

\noindent
The previous analysis giving the asymptotic equilibrium for confined jets can
be pushed one step further to the first order terms in the bending of the lines
(Paper III, Vlahakis \& Tsinganos 1997, henceforth VT97). We also assume there
that the jet becomes asymptotically cylindrical. An expansion  of $G$ and $M$
can be made then to get an idea of the fluctuations that exist far from the
region of the initial acceleration of the wind, 
$$
G^2= G^2_\infty (1 + \vartheta)
\,,
\eqno (\hbox{5.1a})
$$
$$
M^2= M^2_\infty (1 + \mu)
\,,
\eqno (\hbox{5.1b})
$$
where $\vartheta , \, \mu \ll 1$. 
Conversely to  Paper III, we must also expand the pressure as
$$
\Pi= \Pi_\infty (1 + p)
\,.
\eqno (\hbox{5.2})
$$
Thus we obtain the
harmonic oscillator equation for the perturbed jet radius
(see Appendix C for details):
$$
\ddot\vartheta 
+ \left( 2\pi r_* / \Lambda_{\rm osc.}\right)^2 \vartheta = 0
 \,,
\eqno (\hbox{5.3})
$$
where $\Lambda_{\rm osc.}$ is the wavelength of the oscillations.
We can write the wavelength of the oscillations in the form of VT97, Eq. (28), 
namely
$$
\left( 2\pi r_* \over\Lambda_{\rm osc.}\right)^2
={2\lambda^2\over (1-M_\infty^2)^2}
\left[2 + {\mu\over\vartheta}{(2M_\infty^2-1)G_\infty^4-M_\infty^4\over
M^2_\infty(1-M_\infty^2)}\right]
\eqno (\hbox{5.4})
$$
with in the present case 
$$
{\mu \over \vartheta} = 
{\displaystyle 
{\kappa \over \lambda^2}  
  - {G_\infty^2 \over M_\infty^4}
  { M_\infty^4 + G_\infty^4 (1-2M_\infty^2)
  \over 2 (1-M_\infty^2)^2 } 
\over\displaystyle
    {\kappa\over\lambda^2} 
   - G^2_\infty {(1-G_\infty^2)^2\over (1-M_\infty^2)^3} 
}  
\,.
\eqno (\hbox{5.5})
$$ 
Plots of the asymptotic wavelength $\Lambda_{\rm osc.}$ {\it vs.}
$\epsilon/2\lambda^2$ for four representative values of $\Pi_\infty$ 
are shown in Fig. 8. In each of these plots, the values of
$\Lambda_{\rm osc.}$ are plotted for the range of the pressure parameter 
$\kappa/2 \lambda^2$ as in Figs. 1 to 3. 
  
In the domain of magnetocentrifugal jets the wavelength behaviour is very 
similar to the one found in  Paper III as expected.\footnote{In Fig. 3 of 
 Paper III,
the plot of the wavelength in the region where 
$M_\infty$ is close to one has to be corrected, due to the 
presence of a $M^4_\infty$ at the 
denominator of Eq. (5.23) that should be replaced by a $M^2_\infty$. Fortunately this 
change does not affect the curve in the region of astrophysical interest, 
 mainly for $M_\infty \gg 1$.}  
In the case of $\kappa=0$ studied in  Paper III, Eq. (5.4)  takes the very 
simple form (Vlahakis, private communication)
$$
\left( 2\pi r_* / \Lambda_{\rm osc.}\right)^2=
{4\lambda^2\over M_\infty^2 (M_\infty^2-1)}
\eqno (\hbox{5.6})
$$
In the domain of pressure confined jets (which corresponds always to
underpressured jets, $\kappa>0$), the wavelength of the oscillations
behaves like the Alfv\'en number despite of the decrease of the
cylindrical radius. In particular, as the rotator slows down 
($\epsilon/2\lambda^2$ decreasing) the wavelength increases very slowly.
Nevertheless in the limit $\epsilon/2\lambda^2 \rightarrow -\infty$ where
we enter the hydrodynamic regime the wavelength must eventually diverge. The 
similarity of the curves of Figs. 2 and 8 on logarithmic scales, which reflects
the similar behaviour of $M_\infty$ and $\Lambda_{\rm osc}$, can be easily 
understood
in the limit of large values of $M_\infty$ and $V_\infty/V_*$. In this limit 
-- which is by the way expected to be the case for most observed jets -- we 
get that the wavelength is just proportional to the square of the Alfv\'en 
number, $\Lambda_{\rm osc}\sim M^2_\infty\,.$
Note also that, if the wavelength of the oscillations can be related to 
the observed morphology of the jets, we may have here an indirect estimate of 
the magnitude of the poloidal Alfv\'en number. 

For overpressured jets ($\kappa<0$), the behaviour of $\Lambda_{\rm osc.}$
is similarly following the increase of the Alfv\'en number as the pressure
becomes more important, but as it enters the domain of pressure supported jets
after the maximum velocity, the oscillations disappear. In fact in
this last case we can see that the pressure gradient supporting 
the jet cannot restore equilibrium against the confining magnetic pinch.
This could imply that the solutions of this class are unstable. 
 
We must notice finally that oscillations in collimated winds
are quite a general result, not restricted to our class of meridional
self-similar solutions. In fact  oscillating structures have been found not 
only
in other self-similar flows (Chan \& Henriksen 1980, Bacciotti \& Chiuderi 
1992, Del Zanna \& Chiuderi 1996, Contopoulos \& Lovelace 1994,
Contopoulos 1995, VT98), but also in more general analyses of  
axisymmetric outflows (see e.g. Pelletier \& Pudritz 1992).


\section{Asymptotic equilibrium for non collimated flows}

\subsection{Existence of asymptotically non collimated flows}

If the flow is not cylindrical asymptotically, then $F_{\infty}$ must take a
value in the interval $0\le F_\infty < 2$, with $F_{\infty} =0$ corresponding
to conical asymptotics. Assuming that this value of $F_{\infty}$ and the
corresponding shape may be achieved rather slowly, at large distances $R$ from
the central object the analytic expression of $F(R)$ can be written as
$$
F(R \rightarrow \infty) = F_\infty + {c\over \ln R} + 
\sum_{n=1}^{n\rightarrow \infty} 
{c_n \over R^n} 
\,,
\eqno(6.1)
$$
where $c$ and $c_n$ are constants. As $R\longrightarrow \infty$ the
dominant term is the logarithmic one and thus we may keep only this term in the
expansion, such that from Eq. (2.6) and (3.4c) we obtain
$$
G^2(R\rightarrow\infty)={R^{2-F_\infty} \over f_\infty \ln^c(R)} 
\,,
\eqno(6.2\hbox{a})
$$
$$
M^2(R\rightarrow\infty)={R^{2-F_\infty} \over f_\infty \ln^c(R)}
{V_\infty \over V_*}
\,,
\eqno(6.2\hbox{b})
$$
where $f_{\infty}$ is a constant (TTS97,  Paper III). By substituting these
expressions in the definition of the integral $\epsilon$ of Eq. (3.11) and
keeping the dominant terms we get
$$
{\epsilon \over 2\lambda^2}= 
- {1\over 2\lambda^2} {V_\infty^2 \over V_*^2}
\left[\kappa + {1-F_\infty^2/4 \over R^{F_\infty} 
f_\infty \ln^c(R)}\right]
+ {V_*\over V_\infty}
$$
$$
+ {f_\infty \ln^c(R)\over 2 R^{2 - F_\infty}}\left(1 - 
{V_\star \over V_\infty}\right)^2
\,.
\eqno(6.3)
$$
Assuming $\kappa \neq 0$, Eq. (6.3) shows clearly a rather general result: 
{\sl a diverging asymptotic velocity is inconsistent with the constancy 
of $\epsilon$}.
The cylindrical radius and Alfv\'en number may be unbounded, but the terminal
flow speed cannot, no matter what is the exact value of $F_{\infty}$. This is
also in agreement with the fact that the fraction of the heating term converted
to kinetic energy and $\propto \kappa$ cannot diverge (see Sec. 3.1). 

Further insight in the general behaviour of the asymptotics can be gained by
considering the dominant terms in the transverse momentum equation which gives
a relation for the pressure (see Eq. A.4c in Appendix A), 
$$
-{\kappa \Pi (R \rightarrow \infty) \over 2\lambda^2}  
   = f_\infty{V_*^2\over V_\infty^2}{\ln^c(R) \over R^{2-F_\infty} } 
\,.
\eqno(6.4)
$$
The {\it first} conclusion from this relation is that the asymptotic pressure
must vanish because the r.h.s. term (due to the magnetocentrifugal
terms) always vanishes for non cylindrically collimated outflows. 

A {\it second} conclusion is that, {\sl independently of the value 
of $\epsilon$, underpressured flows ($\kappa>0$) must be cylindrically 
collimated}, as found in
Sec. 4. This can be explained taking into account that, when the streamlines
try to expand, then pinching by both  magnetic forces and pressure gradient
will dominate over all other forces (Eq. 6.4). To maintain the forces
equilibrium a strong bending of the lines towards the axis is required,
$R{\hbox{d}F/\hbox{d}R} >0$, such that the system must relax towards a
collimated configuration. Then only overpressured outflows ($\kappa<0$) may be
non collimated. 

\subsection{Asymptotically paraboloidal or radial overpressured flows}

Assuming $\kappa < 0$, we discuss separately the case of paraboloidal
($F_{\infty} > 0$) and radial asymptotics ($F_{\infty} = 0$). 

\subsubsection{Paraboloidal asymptotics, $F_{\infty} > 0$}
Independently on the value of $c$, Eq. (6.3) further simplifies as follows
$$ 
{\epsilon \over 2\lambda^2}= + {|\kappa|\over 2\lambda^2} {V_\infty^2 \over
V_*^2} + {V_*\over V_\infty} 
\,, 
\eqno(6.5) 
$$ 
which implies that {\sl paraboloidal shapes can exist only for $\epsilon>0$}. 
In
other words, only overpressured outflows from an efficient magnetic
rotator may eventually achieve a paraboloidal shape. This can be physically
understood because such a structure implies some collimation that can be
achieved only through magnetic forces in the present case. 

We remark that in the case $\kappa=\epsilon=0$ Eq. (6.5) can be fulfilled
only for diverging asymptotic velocity (Paper III). However, as we have seen before, 
such a case would  require an infinite heating rate, so that such kind of 
solutions should be considered as unphysical.

\subsubsection{Radial asymptotics, $F_{\infty} = 0$}
If $c \neq 0$, Eq. (6.5) and the previous 
remarks still hold. For $c=0$, Eq. (6.3) becomes
$$
{\epsilon \over 2\lambda^2}= 
{1\over 2\lambda^2} {V_\infty^2 \over V_*^2}
\left(|\kappa| - {1 \over  f_\infty}\right)
+ {V_*\over V_\infty}
\,,
\eqno(6.6)
$$
and radial asymptotics can exist {\it a priori} for both negative and positive
values of $\epsilon$. 

This simple analysis shows that overpressured flows ($\kappa<0$) can attain
only radial streamlines if they are IMR ($\epsilon<0$), while they can  have 
cylindrical, paraboloidal or radial asymptotics if they are associated with
an EMR ($\epsilon>0$).

We point out finally a common feature for winds with parabolic or radial
asymptotics. From Eq. (6.4) we see that temperature goes to a constant value
along each fieldline 
$$
T (R\rightarrow\infty, \alpha) \propto
{P\over\rho}={1\over2}V_*^2{1+\kappa\alpha \over 1+\delta\alpha} 
\left(-{2\lambda^2\over\kappa}\right)
{V_*\over V_\infty}
\,,
\eqno(6.7)
$$
i.e. all uncollimated solutions are isothermal asymptotically. This is
consistent with the results of Tsinganos \& Trussoni (1991), who analyzed
solutions with prescribed radial streamlines [$F(R)=0$]. More in general, we 
can
expect that in non collimated outflows some heating is always necessary in the
asymptotic regions. Conversely in a cylindrical jet, where the pressure and 
density are constant, the temperature is also constant but the heating rate
in the flow vanishes (Eq. 3.1b,c). 


\section{Discussion and astrophysical implications}
   
\subsection{Summary of the main results}

We have presented here the asymptotic 
properties of superAlfv\'enic outflows which are self-similar in the 
meridional direction. 
The terminal Alfv\'en number, $M_\infty$, the dimensionless asymptotic
radius of the jet, $G_\infty$ and velocity, $V_\infty/V_*$ depend only 
on three parameters (see Eqs. 4.4). Besides the terminal pressure $\Pi_\infty$, the two other 
crucial parameters are:

\noindent{$\bullet$} \underbar{$\kappa/2  \lambda^2$},  
connecting the radial and longitudinal components of the gradient of the 
gas pressure. Thus, the outflow can be either overpressured ($\kappa <0$),  
or underpressured ($\kappa >0$) with respect to the rotational axis.

\noindent{$\bullet$} \underbar{$\epsilon / 2 \lambda^2$},  
which measures the  magnetic contribution
to the collimation of the outflow. Thus, we may divide the sources of
outflows into two broad classes : {\it Efficient Magnetic Rotators} (EMR) 
corresponding to positive values of $\epsilon$ and a strong magnetic 
contribution to collimation, 
and {\it Inefficient Magnetic Rotators} (IMR) which have negative 
$\epsilon$ and can collimate outflows only with the help of the 
 gas pressure.
\footnote{This does not imply that collimated jets from IMR 
are asymptotically pressure confined. They may be magnetically 
confined at infinity if the terminal pressure $\Pi_\infty$ is very 
small. However in such jets the gas pressure 
\underbar{always} plays a crucial role in the achievement of the 
final collimation through
strong pinching gradients in the intermediate region between the
base and infinity.} 

The absolute criterion for cylindrical collimation is given by the sign 
of a combination of the two parameters $\kappa/2\lambda^2$ 
and $\epsilon/2\lambda^2$. This new parameter

\noindent{$\bullet$} \underbar{$\epsilon^{\prime}/ 2 \lambda^2$} 
is related to the variation across the
 streamlines of the various energy contributions which govern the flow 
dynamics (magnetocentrifugal, thermal, etc). Thus, a positive value of
$\epsilon^{\prime}$ provides cylindrical asymptotics while negative values
are required for having uncollimated flows.

\begin{figure}
\epsfig{figure=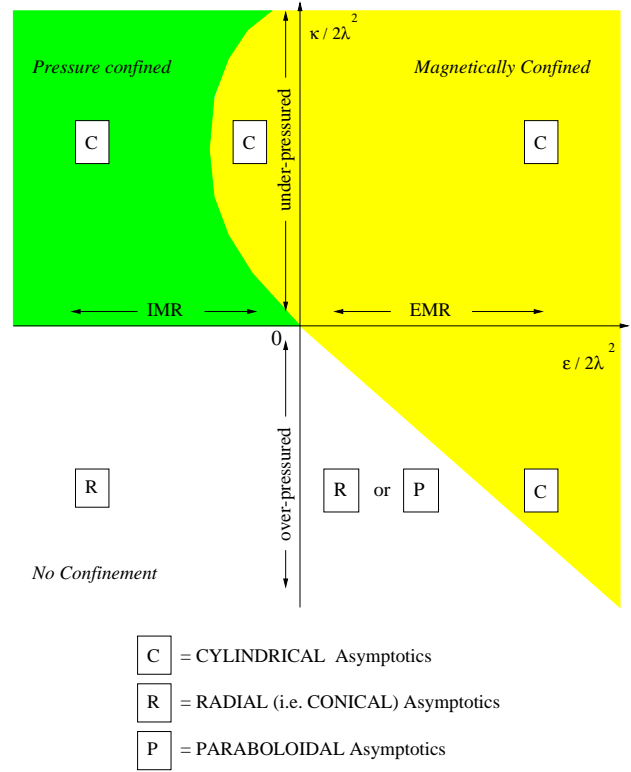}
\caption{A sketch of the various confinement regimes and asymptotical shapes 
of the flow in the plane [$\kappa/2\lambda^2,\,\epsilon/2\lambda^2$].
The pressure confined regime corresponds to the area filled with dark 
grey while the magnetically confined regime corresponds to the light
grey filled area. In the remaining area only uncollimated solutions are
 found.
}
\end{figure}

Despite this simple criterion, the asymptotic behaviour of the outflow still
depends on the value of each parameter taken separately as shown in Fig. 9
[see also Figs. (1--3)].

\noindent{$\bullet$}{\underbar{Underpressured outflows}}. 
For $\kappa > 0$ the wind always obtains
cylindrical asymptotics. The jet is supported by the centrifugal force,
while collimation can be provided either by the magnetic pinching or
the gas pressure, depending on $\epsilon$ and the value of the 
asymptotic gas pressure $\Pi_{\infty}$. We repeat however that for 
IMR the state of asymptotic magnetic confinement can be achieved only 
through the strong pressure gradients occurring between the base of the flow 
and infinity. If $\Pi_{\infty}$ does not vanish, the jet radius
has a maximum when moving from the magnetic to the thermal regime 
(by reducing $\epsilon/ 2 \lambda^2$). 

\noindent{$\bullet$}{\underbar{Overpressured outflows}}. 
When $\kappa < 0$ the jet can be confined
by the magnetic pinch only, and is supported either by the centrifugal force or
by the thermal pressure gradient. If $\Pi_{\infty}$ does not vanish, the jet 
terminal velocity has a maximum when moving from the centrifugal to the 
thermal regime. Moreover, in overpressured outflows, cylindrical 
configurations are attained only for values of $\epsilon/2 \lambda^2$ higher 
than some threshold depending on the pressure parameters, 
$\Pi_{\infty}$ and $\kappa$. Below this positive threshold value the outflow
reaches conical or paraboloidal asymptotics if $\epsilon>0$ (EMR) or
only purely conical asymptotics if $\epsilon<0$ (IMR).

Cylindrical collimation seems to be quite a natural end state 
for superAlfv\'enic outflows with non vanishing asymptotic pressure,
as also found by other studies based on the radial self-similar approach (Li
1995, 1996, Ferreira 1997, Ostriker 1997), or on the full asymptotic treatment
of the MHD equations (Heyvaerts \& Norman 1989). However in the present case
the collimation can be provided not only by the magnetic pinch, but also
by the thermal pressure gradient. This is consistent with our self-similar 
scenario,
suitable to model winds close to their rotational axis, where the thermal
effects are essential to drive the outflow. We also point out that our results
are consistent with those of TTS97, where again different collimation regimes
can be found. 
Finally, we remark that cylindrically collimated streamlines  most of times 
undergo oscillations with different wavelengths.

\subsection{Astrophysical application}

As in  Paper III and TTS97, the present results could be particularly suitable
to model the physical properties of collimated outflows associated with 
Young Stellar Objects (YSO). However, since here we have analyzed only the 
asymptotic properties of winds, we
will shortly  discuss only a simple possible scenario based on the physical
difference between EMR and IMR. 

Let us consider a rapidly rotating magnetized protostar at the beginning of its
evolution. In such conditions this object could be considered as an EMR, with a
well collimated, magnetically confined jet. 
At the same time the pressure inhomogeneity $|\kappa|$ and the asymptotic 
pressure may take rather high values due to the inhomogeneous and anisotropic 
environment in which the jet is found.  
In the early phases of stellar evolution the outflow extracts quite 
efficiently angular momentum from the protostar, reducing its spinning rate. 
From the point of view of our model this means that the system moves from 
the state of an EMR to that of an IMR as it lowers the value of $\epsilon$. 
Of course the details of the
evolution may be more complicated due to the feeding of the wind by the 
surrounding accretion disk. Nevertheless the net end result should be a 
decrease of the spinning rate and subsequently of the efficiency of the 
magnetic rotator (Bouvier et al. 1997). 

If the jet is initially underpressured, for example being embedded in 
a dense molecular cloud, as it approaches the regime of the IMR a gradual  
widening of its radius is expected. This widening may reach a maximum 
value, wherein the magnetic and thermal contributions to the confinement 
are comparable. At this stage it is reasonable to expect a decrease of 
the asymptotic pressure because of this widening. At the same time, 
we may have a more homogeneous flow, i.e., effectively a decrease of 
$\kappa$. Thus, in the subsequent evolution of the central young star, we 
expect that the jet continues increasing its radius. 
According to this scenario, we can imagine an evolutionary
track of the outflow along the maxima from  Figs. 1d to Fig. 1b, 
wherein there is an equal contribution by the thermal and magnetic forces 
confining the plasma. 
In this sequence, the terminal velocity always increases, Figs. 3d to 3b, as
$\kappa$ and the asymptotic pressure $\Pi_\infty$ get lower and lower values.  
Finally, the outflow becomes a loosely collimated wind.  

In the above scenario, it is essential to have a decrease of $\kappa$ and 
$\Pi_\infty$ during the evolution. Because, otherwise, we see that an IMR
($\epsilon<0$) together with a strongly underpressured plasma 
(high $\kappa \Pi_\infty$) have the result to over collimate the wind with 
an asymptotic radius comparable to its Alfv\'en radius. 
This result would be somehow in contradiction with the observed
radii of outflows from YSOs which are expected to be much larger than the 
Alfv\'en radius. Instead, we propose that both thermal and 
magnetic confinement decrease simultaneously during the
evolution of the central source. 

However, jets from planetary nebulae (PN) may
present a totally different situation where the primary source of confinement 
is a strong pressure gradient, $\kappa\Pi_\infty\gg 0$, associated with a 
source which is a very inefficient magnetic rotator, although with a non 
zero magnetic field. The terminal radii of jets from PN are indeed observed 
to be rather small after some huge initial widening (Frank 1998).
 We also note that our analysis favours a hydrodynamical origin of 
jets from PN similarly to the GWBB model (cf. Mellema \& Frank 1997, Frank 
1998) and contrarily to a pure magnetic origin of the refocusing of the wind.

The analysis of overpressured jets is more complex at first glance as 
different regimes are possible for the same $\epsilon$. 
Let us first consider that the jet is initially quite narrow,
centrifugally supported and originates from an EMR; i.e., it is on the lower 
branches of the thin grey lines ($\kappa<0$) of Figs. 1--3. 
As in the previous case, the
rotation slows down ($\epsilon$ decreases), the outflow rapidly opens and the
velocity rises. However, below a threshold value of $\epsilon/ 2 \lambda^2$ the
flow becomes uncollimated (see Fig. 9). 

If, conversely, at the beginning the jet is pressure supported (i.e., on 
the upper branches of the dotted lines in Figs. 1b--d, 2b--d),  the behaviour 
is quite ambiguous: a reduction or an increase of the 
jet radius and velocity critically depends on the initial conditions 
($|\kappa|$, asymptotic pressure, jet radius) 
when the star starts slowing down its rotational 
rate. Then it is difficult to model the possible
evolution of a pressure supported jet. But, as we said previously, the absence
of oscillations in this region may indicate that such equilibria are in fact 
unstable and never attained practically.

It is evident from the above that the possible outflow evolution is 
critically related to its physical conditions, namely if it is underpressured
or overpressured. In which of these regimes the outflow can be found depends
on the detailed history of the wind. We remind that 
the thermodynamic conditions across the jet in its asymptotic regime
depend also on the assumed structure of density ($\delta$) and the 
intensity of the gravitational field ($\nu$). These two parameters
do not enter in the present analysis, but are essential for the 
thermal acceleration of the wind (see  Paper III and TTS97).
Therefore the next step in the present analysis is to make a careful
parametric study of the numerical solutions, solving the 
set of Eqs. (A.4). This is also demanded in order to make a
detailed comparison with observational data and implies that we construct
solutions that connect the
base of the flow with the superAlfv\'enic region fulfilling the regularity
conditions at the critical points (Sauty et al., in
preparation). In particular, it will be crucial to see whether or not all the
asymptotic regimes presented here can be attained. 

\subsection{Future directions of study}

The present results have been obtained in the framework of a 
self-similar treatment of the axisymmetric MHD equations. This implies some `a
priori' constraints on the structure of the solutions, that we summarize in the
following. 

The surfaces with the same Alfv\'enic number are spherical
 [$M(R, \theta) \equiv
M(R)$], with the velocity vanishing on the equatorial plane. Furthermore the
$\theta-$ and $r-$ components of the gradient of the gas pressure are linearly
related. These assumptions are not too constraining if we consider our
model as suitable to describe the physical properties of  the flow around
the rotational axis. We remind also that our results are consistent with those
found in TTS97, where the two components of $\nabla P$ were unrelated in a wind
with prescribed cylindrical asymptotics.

Such limits of the present treatment could anyway be overcome by a
different scaling of the physical variables with the colatitude. It has been 
shown in VT98 that the assumptions of Eqs.  (2.3)
are just a particular case of a more general class of solutions, with no 
vanishing velocity on the equator and with a more complex expression for 
the pressure (another particular case is the one studied in Lima et al. 
1996; see also Vlahakis 1998).
In such a case the set of the MHD equations lead to a closed system whose
treatment does not require any further `a priori' assumption as the
relation between the components of the pressure gradient (as in Paper III
 and in the present study) or the prescription of the streamline 
 shape (as in TT97).

Finally, we point out once more that, contrary to the radial self-similar
studies, the role of the thermal structure of the
flow, which is related to the details of processes of input/output of heating 
in the gas, is essential in our model and not only in accelerating the flow
but also in constructing its global shape. The problem of the energetic
 behaviour of astrophysical plasmas in
different astrophysical contexts (solar and stellar coronae, YSO, AGN) is
however still open. 


\begin{acknowledgements}

We are indebted to Dr. N. Vlahakis for stimulating discussions on
the course of this work and Dr. F.P. Pijpers for carefully reading and 
commenting the manuscript. 
C.S. and K.T. acknowledge financial support from
the French Foreign office and the Greek Ministery of Research and Technology. 
E.T. thanks the Observatoire
de Paris and the Physics Department of the University of Crete
for hospitality, and the Agenzia Spaziale Italiana for financial
support.

\end{acknowledgements}
\appendix


\section{Appendix A}

The classical equations of ideal MHD steady flows are
$$
\nabla \cdot \vec{B} = \nabla \cdot (\rho \vec{V}) = \nabla \times (\vec{V} 
\times \vec{B}) = 0
\,, 
\eqno(\hbox{A.1a})
$$
$$
\rho (\vec{V} \cdot \nabla) \vec{V} - (\vec{B} \cdot \nabla) \vec{B} /4 \pi 
$$
$$=
- \nabla (p + B^2/8 \pi) - {\cal G} {\rho {\cal{M}} \over r^2}
{\bf e}_r
\,,
\eqno(\hbox{A.1b})
 $$
$$
\rho \vec{V} \cdot \left (\nabla h - {{\nabla P} \over {\rho}} 
\right )=q\,,
\eqno(\hbox{A.1c})
$$
where $h$ is the enthalpy of the perfect gas, $q$ is the local volumetric 
heating rate including true heating and cooling, ${\cal G}$ is the gravitational 
constant, ${\cal M}$ is the mass of the central object and the other symbols 
have their usual meaning.

Under the assumption of steady state and axisymmetry, the existence of 
free integrals as defined in the main text gives the usual forms for the 
poloidal ($p$) components of
the magnetic field $\vec{B}$ and the velocity $\vec{V}$, using spherical ($r, 
\theta, \varphi$) or cylindrical ($\varpi, \varphi, z$) coordinates
(for details see Tsinganos 1982)
$$
\vec{B}_p= {\nabla A \over \varpi} \times 
{\bf e}_\varphi\,, 
\qquad
\vec{V}_p = {\Psi_A \over 4 \pi \rho} \vec{B}_p\,,
\eqno(\hbox{A.2a})
$$
\noindent
while the toroidal components are
$$
B_{\varphi} = - {{L(A) \Psi_A} \over \varpi} 
{{1 - \varpi^2 \, \Omega(A)/L(A)} \over {1 - M^2}}
\,,
\eqno(\hbox{A.2b})
$$
$$
V_{\varphi} = {L(A)  \over \varpi} \,
{\varpi^2 \, \Omega(A)/L(A) -M^2 \over 1 - M^2}
\,,
\eqno(\hbox{A.2c})
$$
\noindent
where $M$ is the poloidal Alfv\'en Mach number as defined in Eq. (2.1).

Considering our assumptions, Eqs. (2.3), we get that the components of 
the velocity  and magnetic field reduce in our model to
$$
B_r =B_{*} {1\over G^2(R)}\cos\theta
\,,
\eqno (\hbox {A.3a})
$$
$$
B_\theta =-B_{*} {1\over G^2(R)}{F(R)\over 2}\sin\theta
\,,
\eqno (\hbox {A.3b})
$$
$$
B_\varphi = - B_{*} {\lambda \over G^2(R)}
{\displaystyle 1 - G^2(R) \over 1 - M^2(R) }{R\sin\theta}
\,,
\eqno (\hbox {A.3c})
$$
$$
V_r = V_{*}  {M^2(R)\over G^2(R)} { \cos\theta \over
\sqrt{1+\delta \alpha(R,\theta)}  }
\,,
\eqno (\hbox {A.3d})
$$
$$
V_\theta =-V_{*} {M^2(R)\over G^2(R)}{F(R)\over 2}
{  \sin\theta \over \sqrt{1+\delta \alpha(R,\theta)}  }
\,,
\eqno (\hbox {A.3e})
$$
$$
V_\varphi = V_{*}  {\lambda \over G^2(R)}
{ G^2(R) - M^2(R) \over 1- M^2(R)}
{R\sin\theta \over \sqrt{1+ \delta \alpha(R,\theta) } }
\,.
\eqno (\hbox {A.3f})
$$
These last equations, together with Eqs. (2.3), can be combined with 
the poloidal components of the momentum equation, Eq. (A.1b), to give three 
independent equations for four unknowns $\Pi(R)$,
$F(R)$, $G(R)$ and $M^2(R)$.  The system is closed with Eq. (2.6) for $G$.
Two of these equations  arise from momentum-balance in the radial direction,
while  the third one from momentum-balance in the meridional direction. 
Thus, we obtain 
$$
{\hbox {d} \Pi \over \hbox {d} R}  
+  {2 \over G^4 }
   \left[ {\hbox{d} M^2 \over \hbox{d} R} + {M^2 \over R^2} (F-2) \right]
+  {\nu^2 \over M^2 R^2 } =0 
\,,
\eqno({\hbox {A.4a}})
$$
$$
-\kappa FR\Pi -\kappa R^2 { \hbox {d} \Pi \over \hbox {d} R} 
+{2 \over G^2} {\hbox {d} M^2 \over \hbox {d} R}
- {\delta \nu^2 \over M^2 G^2}
$$
$$
- { F \over 2 R G^2} \left[ R{\hbox{d}F\over\hbox{d}R} +F^2-F-2\right]
+ {M^2\over R G^2}\left[ {F^2\over 2} +F-4\right]
$$
$$
+ {2\lambda^2 R\over G^2 M^2}{(M^2 - G^2)^2 \over (1 - M^2)^2}
- {2\lambda^2 R\over G^2}{1 - G^2 \over (1 - M^2)^2}(F-2G^2)
$$
$$
- {2\lambda^2 R^2\over G^2}{(1 - G^2)^2 \over (1 - M^2)^3}
                           {\hbox {d} M^2 \over \hbox {d} R}
=0
\,,
\eqno({\hbox {A.4b}})
$$
$$
-{F\over 2}R{\hbox {d} M^2 \over\hbox {d} R}
+ \kappa R^2 G^2 \Pi 
+ {1\over 2} \left[ R{\hbox{d}F\over\hbox{d}R} +F^2-F-2\right]
$$
$$
- {M^2 \over 2} \left[ R{\hbox{d}F\over\hbox{d}R} + {F^2\over 2}-F\right]
$$
$$
- { \lambda^2 R^2 \over (1 - M^2)^2 } 
   \left[ {(M^2 - G^2)^2 \over M^2}- 2 (1 - G^2)^2 \right]
=0
\,.
\eqno({\hbox {A.4c}})
$$
In Sec. 5 (oscillations of cylindrical jets), the first order expansion 
scheme of the previous momentum equations amounts to saying that we have the
 following expression for the force balance across the fieldlines 
$$
\rho (\vec{V}_p.\nabla)\vec{V}_p 
-{1\over4\pi}(\nabla\times\vec{B}_p)\times\vec{B}_p
$$
$$
= 
{B_\varphi\over 4\pi\varpi}\nabla(\varpi B_\varphi)
  - {\rho V_\varphi^2\over\varpi}\nabla \varpi - \nabla P
\,.
\eqno (\hbox{A.5})
$$


\section{Appendix B}

\subsection{On the variation of the specific energy}

In Eq. (3.3b), we find successively five terms which correspond to 
the {\bf variation}, in units of the volumetric energy, of the 
magnetic rotator between any streamline ($\alpha$) and the polar axis (pole)
of 

\noindent
(i) the poloidal kinetic energy, 
$$
{M^4\over R^2 G^2}\left[{F^2\over 4} - 1 \right]
= {2\lambda^2 \over \rho(R,\alpha)L(\alpha)\Omega(\alpha)} \times
$$
$$
\times \left[ {1\over 2}\rho(R, \alpha) V^2_p(R,\alpha) 
-  {1\over 2}\rho(R, {\rm pole}) V^2_p(R,{\rm pole}) \right]
\,,
\eqno (\hbox{B.1})
$$
(ii) the gravitational energy, 
$$
 -{\delta\,\nu^2 \over R}
=  -{2\lambda^2 \over \rho(R,\alpha)L(\alpha)\Omega(\alpha)} 
{{\cal G}{\cal M} \over R}
\left[ \rho(R,\alpha ) - \rho(R,{\rm pole}) \right]
\,,
\eqno (\hbox{B.2})
$$
(iii) the azimuthal kinetic energy ($\equiv 0$ along the polar axis), 
$$
{\lambda^2 \over  G^2}
  \left[ M^2 - G^2 \over 1 -M^2\right]^2
={2\lambda^2 \over \rho(R,\alpha)L(\alpha)\Omega(\alpha)} 
\left [{1\over 2}\rho(R,\alpha) V_{\varphi}^2(R,\alpha) \right ]
\,,
\eqno (\hbox{B.3})
$$
(iv) the Poynting flux ($\equiv 0$ along the polar axis), 
$$
2\lambda^2\left[1 - G^2\over 1 - M^2 \right]
=-{2\lambda^2 \over \rho(R,\alpha)L(\alpha)\Omega(\alpha)} 
\left [ {\Omega(\alpha) \, \over \Psi_A(\alpha)} \varpi B_\varphi(R,\alpha)
\right ]
$$
$$
={2\lambda^2 \over \rho(R,\alpha)L(\alpha)\Omega(\alpha)} 
\left[L(\alpha)\Omega(\alpha) - \varpi^2(R,\alpha)\Omega^2(\alpha) \right] 
\,,
\eqno (\hbox{B.4})
$$
(v) the thermal content, 
$$
\kappa\left[{\Gamma\over \Gamma - 1} \Pi M^2
        -\int_{R_o}^{R}{\cal Q}(R)dR \right]
={2\lambda^2 \over \rho(R,\alpha)L(\alpha)\Omega(\alpha)} 
$$
$$
\times  \Bigg\{ \rho(R,\alpha) 
[h(R,\alpha) - \Theta_{R_o}^R(\alpha) ]
$$
$$
- \rho(R,{\rm pole})  [h(R,{\rm pole})  
 - \Theta_{R_o}^R({\rm pole}) ]  \Bigg\}
.
\eqno (\hbox{B.5})
$$
In order to write Eqs. (3.7), we simply calculate the previous terms (B.1-B.5)
at the base of the flow $r_o$
$$
\Delta {\cal E}
= 
-{\delta\,\nu^2 \over R_o}
+ {\lambda^2\over G_o^2}
  \left[ M_o^2 - G_o^2 \over 1 -M_o^2\right]^2
$$
$$
+ 2\lambda^2 \left[1 - G_o^2\over 1 - M_o^2 \right]
 +\kappa\left[{\Gamma\over \Gamma - 1} \Pi_o M_o^2\right]
\,.
\eqno(\hbox{B.6})
$$
There the poloidal kinetic energy is negligible and consequently the term in
Eq. (B.1) is zero. Moreover the thermal content reduces to the enthalpy such
that in Eq. (B.5) only the enthalpic terms remain. Combining Eqs. (B.2) to
(B.6), we get Eqs. (3.7). 

\subsection{On the Energetic definition of EMR and IMR}

Now using Eqs. (3.2) and (3.10) we may write
$$
E(\alpha)={1\over 2}V_*^2
     {{\cal E}(1 + \kappa \alpha) + \alpha  \epsilon \over 1+\delta\alpha }
\,,
\eqno (\hbox{B.7})
$$      
where we see, as stated in the main text, that $\epsilon$ is the transverse 
variation of the volumetric energy once we have removed the thermal terms 
that linearly scale with factor $\kappa$.
Moreover note that from Eq. (3.4a) we get 
$$
{{\cal E}\over 2\lambda^2}= 
{ \rho_o({\rm pole}) [ h_o({\rm pole})  + E_{\rm G,o}   ]
\over \rho_o(\alpha)E_{\rm {MR}}
}
\,,
\eqno (\hbox{B.8})
$$
which is the pending expression to Eq. (3.7a).
Noting that 
$$
\kappa = {\Delta(\rho h)\over \rho({\rm pole})h({\rm pole})} 
= {\Delta P \over P({\rm pole})}
\,,
\eqno (\hbox{B.9})
$$ 
we combine Eqs. (3.7a), (B.8) and (3.10) to get
$$
{\epsilon  \over 2\lambda^2} = 
{ \Delta \left[ \rho_o (
E_{\rm {Poynt., o}} + E_{{\rm R},o} )\right] 
\over \rho_o(\alpha)E_{\rm {MR}} }
$$
$$
+{ E_{\rm G,o}   \over E_{\rm {MR}}}{\rho_o({\rm pole})\over \rho_o(\alpha)}
\left[ {\Delta \rho_o\over \rho_o({\rm pole})} - {\Delta P_o\over P_o({\rm pole})}
\right]
\,.
\eqno (\hbox{B.10})
$$
The first term of the r.h.s. of this last equation simplifies to the two first
terms in the numerator of Eq. (3.12a) as the Poynting flux and the rotational 
energy vanish along the pole. The second term of the r.h.s. of Eq. (B.10) 
can be rewritten to give $\Delta E_G^*$ in Eq. (3.12a). In the form presented 
in Eq. (B.10) it appears
how the relative increase of the weight of the plasma can be partially 
compensated by the increasing of the thermal pressure gradient. In this form 
there is no contradiction with the use of the symbol $\Delta$. Conversely,
the  equivalent expression used in Eq. (3.12b) may appear confusing if one
does not remember that this is in fact the variation across the lines of the 
gravitational energy that is not compensated by some thermal driving. 
Nevertheless,
we prefer this last form for its compactness and because it emphasizes the 
role of the temperature.


\section{Appendix C}

From Eqs. (5.1) and (5.2) we may expand $F$ to first 
order in Eq. (2.6),
$$
F  =  2 - R\dot\vartheta
\,,
\eqno (\hbox{C.1a})
$$
while the derivative of $F$ can be also expanded at large $R$ as
$$
{{\rm d}F\over{\rm d}R} = - \dot\vartheta - R\ddot\vartheta
 \approx - R^2\ddot\vartheta
\,.
\eqno (\hbox{C.1b})
$$
From these equations we can expand the momentum equations given in Appendix 
A (Eqs. A.4) replacing the second one by the definition of $\epsilon$.
We still assume in this Section that the flow is asymptotically 
cylindrically collimated. Thus we already have calculated the zeroth order 
equilibrium in
Sec. 4. We know that the asymptotic quantities in the flow are
uniquely determined by the values of $\epsilon/2\lambda^2$, 
$\kappa/2\lambda^2$ and $\Pi_\infty$. The first order terms in the 
transverse momentum equation (see also Eq. A.5) give 
a relation between $\ddot\vartheta$, $\vartheta$ and $\mu$
$$
\ddot\vartheta=
2\kappa\Pi_\infty {G^2_\infty \over 1-M_\infty^2 }
\left[ p+\vartheta - {2 M^2_\infty \over 1-M_\infty^2}\mu \right]
$$
$$
-2\lambda^2 \left[
{2G_\infty^2\over (1-M_\infty^2)^2}\left({1-G_\infty^2\over 1-M_\infty^2}
+{G_\infty^2\over M_\infty^2}\right)\vartheta \right.
$$
$$
\left.
 +{M_\infty^4-G_\infty^4\over (1-M_\infty^2)^3 M_\infty^2}\mu
\right]
\,. \eqno(\hbox{C.2})
$$

This can be combined with Eq. (3.11) that we also expand to
first order -- where again the zeroth order is Eq. (4.4b) -- to  get
a second relation between $\mu$ and $\vartheta$,
$$
\mu\left[- {\kappa\over\lambda^2} 
      + G^2_\infty {(1-G_\infty^2)^2\over (1-M_\infty^2)^3} \right]
$$
$$
+ \vartheta\left[ {\kappa \over \lambda^2}  
               - {G_\infty^2 \over M_\infty^4}
                 { M_\infty^4 + G_\infty^4 (1-2M_\infty^2)
                  \over 2 (1-M_\infty^2)^2 } 
          \right]
=0
\,,
\eqno (\hbox{C.3})
$$
which is identical to Eq. (5.5).
Expanding Eq. (A.4a) we have a relation between the derivatives of 
$\theta$, $\mu$ and $p$, 
$$
\dot p+2{M_\infty^2\over G_\infty^4\Pi_\infty}(\dot\mu-\dot\vartheta)=0
\,,
\eqno (\hbox{C.4a})
$$
that can be integrated to give $p$ as a function of 
$\theta$ and $\mu$, assuming a vanishing constant of integration
$$
p+2{M_\infty^2\over G_\infty^4\Pi_\infty}(\mu-\vartheta)=0
\,.
\eqno (\hbox{C.4b})
$$
By eliminating $\mu$ and $p$ in Eq. (C.2) using Eqs. (C.3)-(C.4b),
we obtain Eqs. (5.3) and (5.4) in Sec. 5.
 

\end{document}